\documentclass[letterpaper,11pt]{article}
\pdfoutput=1 
\usepackage[utf8]{inputenc}
\usepackage{jheppub} 
\usepackage{amsmath}
\usepackage{bbold}
\usepackage{xcolor}
\usepackage{subfig}
\usepackage{graphicx}
\usepackage{hyperref}
\usepackage{mathtools}
\usepackage{jheppub}
\usepackage{amsfonts}
\usepackage{amssymb}
\usepackage{amsthm}
\usepackage{physics}
\usepackage{caption}
\usepackage{comment}
\usepackage{mathrsfs}
\usepackage{lscape}
\newcommand{\widesim}[2][1.5]{
  \mathrel{\overset{#2}{\scalebox{#1}[1]{$\sim$}}}}
\usepackage[T1]{fontenc} 

\title{On unitarity of the Coon amplitude}
\author[a]{Rishabh Bhardwaj,}
\author[a]{Shounak De,}
\author[a,b]{Marcus Spradlin}
\author[a]{and Anastasia Volovich}

\affiliation[a]{Department of Physics, Brown University,\\
	182 Hope Street, Providence, RI 02912, U.S.A.}
\affiliation[b]{Brown Theoretical Physics Center, Brown University,\\
	340 Brook Street, Providence, RI 02912, U.S.A.}

\emailAdd{rishabh\_bhardwaj@brown.edu}
\emailAdd{shounak\_de@brown.edu}
\emailAdd{marcus\_spradlin@brown.edu}
\emailAdd{anastasia\_volovich@brown.edu}

\abstract{The Coon amplitude is a one-parameter deformation of the Veneziano amplitude. We explore the unitarity of the Coon amplitude through its partial wave expansion using tools from $q$-calculus. Our analysis establishes manifest positivity on the leading and sub-leading Regge trajectories in arbitrary spacetime dimensions $D$, while revealing a violation of unitarity in a certain region of $(q,D)$ parameter space starting at the sub-sub-leading Regge order. A combination of numerical studies and analytic arguments allows us to argue for the manifest positivity of the partial wave coefficients in fixed spin and Regge asymptotics.}

\begin{document} 
\maketitle

\newpage

\section{Introduction}
\label{sec:intro}
A four-point amplitude that: i) is crossing symmetric in $s \leftrightarrow t$, ii) has polynomial residues, and iii) is meromorphic (with simple poles only), was first introduced by Veneziano in 1968~\cite{Veneziano:1968yb}\footnote{Here the parameter $\alpha_0 = 0$ for type-I superstrings and $\alpha_0=1$ for bosonic strings. In some papers on this subject, $\alpha_0$ is used interchangeably with $m^2$.}:
\begin{equation}
    \mathcal{A}^V(s,t) = -(s+t)^{\alpha_0-1} \frac{\Gamma(-\alpha_0-s)\Gamma(-\alpha_0-t)}{\Gamma(-2\alpha_0-s-t)}~.
    \label{venezianoamp}
\end{equation}
Its worldsheet realization as a dual resonance model in an operator formalism was established later \cite{susskindhadron,Goto:1971ce,Nambu:1974zg}. Owing to an infinite product representation of the gamma function, the simple pole structure of the Veneziano amplitude is made manifest by the expression:
\begin{equation}
     \mathcal{A}^V(s,t) = \frac{1}{st}\prod_{n=1}^{\infty}\frac{1-\frac{s+t}{n}}{\left(1-\frac{s}{n}\right)\left(1-\frac{t}{n}\right)}~,
     \label{venezianoprod}
\end{equation}
which holds when $\alpha_0=0$. 

A year later, an investigation into the uniqueness of the Veneziano amplitude led Coon to suggest a more generic infinite product representation albeit with logarithmic Regge trajectories \cite{Coon:1969yw}\footnote{It is worth noting that this representation of the amplitude was not first adopted in \cite{Coon:1969yw}, but originally appeared in later papers by the same author \cite{Coon:1972qz}. More recently, beginning with the work of \cite{Figueroa:2022onw}, it has become conventional to adopt this infinite product representation.}:
\begin{equation}
    \mathcal{A}^C(s,t) = (q-1)\exp\left(\frac{\log{\sigma}\log{\tau}}{\log{q}}\right)\prod_{n=0}^{\infty}\frac{\left(1-\frac{q^n}{\sigma\tau}\right)(1-q^{n+1})}{\left(1-\frac{q^n}{\sigma}\right)\left(1-\frac{q^n}{\tau}\right)}~,
    \label{coonkakhoon}
\end{equation}
where
\begin{equation}
    \sigma = 1-(1-q)(\alpha_0+s)~,\quad\quad\tau = 1-(1-q)(\alpha_0+t)~,
\end{equation}
and $q \in [0,1]$ is an arbitrary parameter\footnote{In principle, the deformation parameter $q$ could be any real number, but it is easy to see that there is no chance for the amplitude to be unitarity if $q$ is outside $[0,1]$.}.

In this paper we focus on the Coon amplitude, which smoothly interpolates between the Veneziano amplitude as $q\to 1$ and scalar field theory as $q\to 0$:
\begin{equation}
    \mathcal{A}^{C}(s,t) \overset{q\to0}{\longrightarrow}\frac{1}{s+\alpha_0}+\frac{1}{t+\alpha_0}+1~.
\end{equation}
Following its inception, many aspects of the Coon amplitude have been studied in some detail, including its $n$-point generalization \cite{Baker:1970vxk}, prescriptions for factorization \cite{yu1972first,baker1972twist}, a loop-order extension \cite{Baker:1971jr}, and its operator formulations \cite{Arik:1973eb,Coon:1972te,Chaichian:1992hr,Jenkovszky:1994qg}\footnote{We refer the interested reader to \cite{Romans:1988qs,Romans:1989di} for an early review on the subject.}. However, the physical origin of the Coon amplitude is not yet clear, and a worldsheet realization of the amplitude remains elusive (see \cite{Bernard:1989jq} for attempts in this direction by $q$-deforming the worldsheet conformal group). One of its key physical features is the appearance of an infinite tower of massive states in the form an accumulation point spectrum (similar to that in the hydrogen atom); such features have also been observed in several recent studies in different physical scenarios \cite{Bern:2021ppb, Caron-Huot:2016icg}. More recently, the authors of \cite{Maldacena:2022ckr} have attempted to reproduce such accumulation-like aspects of the Coon amplitude by virtue of open string scattering on D-branes in an AdS background. Also, multi-parameter spaces of four-point amplitudes satisfying the above constraints i)-iii) have been introduced in \cite{Cheung:2022mkw,Geiser:2022exp} with further modified spin-dependent Regge trajectories.

A necessary condition for the Coon amplitude to have physical relevance is that it must be unitary. An early attempt to analyze unitarity was carried out in \cite{Fairlie:1994ad} where it was claimed that the Coon amplitude is \textit{ghost-free}. A more careful and detailed analysis on this issue was carried out more recently in \cite{Figueroa:2022onw}. It was demonstrated by numerical methods that there exists a curve in $(q,D)$ space separating regions where the Coon amplitude does or does not satisfy unitarity in $D$ (spacetime) dimensions. They also analytically show the existence of a certain critical value $q=q_{\infty}(-\alpha_0)$ where the amplitude switches from being unitary in all dimensions to having a critical dimension. However, an  analytic analysis of the unitarity of the Coon amplitude for all $(q,D)$ is still an outstanding problem. In this paper, we study the positivity of the Coon amplitude's partial wave expansion coefficients, in a spirit similar to the recent techniques developed for the Veneziano amplitude in \cite{Arkani-Hamed:2022gsa}. In doing so, we make small strides towards an analytic analysis of Coon unitarity. We achieve this by utilizing several elements of $q$-calculus which have been well studied in the literature \cite{kac2002quantum,ernst2012comprehensive}.

The paper is organized as follows. In section \ref{sec:contour-repn}, we derive a novel contour integral representation for the partial wave expansion coefficients $B^{(D)}_{n,j}(q)$. In section \ref{sec:reggetrajectories} we analyze Regge trajectories (leading and beyond), finding that the leading and sub-leading Regge trajectories are manifestly unitary, while unitarity fails for a certain region in $(q,D)$ space for the sub-sub-leading coefficient $B_{3,0}^{(D)}(q)$. Finally, in section \ref{sec: smallqdeform}, we strive towards obtaining a double-contour integral representation for the coefficients in the near Veneziano case ($1 - q \ll 1$) and apply our formula to study their asymptotic behavior. We conclude with a discussion of our results and possible future directions in section \ref{sec:outlook}. 

\section{A contour integral representation for partial wave coefficients}
\label{sec:contour-repn}
In this section, we establish a contour integral representation for the partial wave expansion of Coon amplitude. The analysis here closely follows a $q$-generalization of the derivation presented in section 3.1 of \cite{Arkani-Hamed:2022gsa}. 

\subsection{A contour representation}

Our starting point is the representation of $\mathcal{A}^\textrm{C}(s,t)$ in terms of the $q$-gamma function\footnote{Our normalization matches the one stated in recent literature \cite{Geiser:2022icl}.}:
\begin{align}
\mathcal{A}^\textrm{C}(s,t) &= q^{\alpha_q(s) \alpha_q(t) - \alpha_q(s) - \alpha_q(t)} \frac{\Gamma_q(-\alpha_q(s)) \Gamma_q(-\alpha_q(t))}{\Gamma_q(1-\alpha_q(s)-\alpha_q(t))}~. 
\label{coonq-gamma} 
\end{align}
This expression can also be recast as a $q$-beta function.  Utilizing its integral representation (\ref{integralq-beta}) (for more details, refer to Appendix \ref{sec:q-gammabeta}) one can rewrite the amplitude as a ``Koba-Nielsen'' like integral \cite{Romans:1988qs}:
\begin{align}
\mathcal{A}^\textrm{C}(s,t) = \frac{q^{\alpha_q(s) \alpha_q(t) - \alpha_q(s) - \alpha_q(t)}}{[-\alpha_q(s)]_q} \int_{0}^{1} d_qz \, \, z^{-\alpha_q(s)} (1-qz)_q^{-\alpha_q(t)-1}~.
\label{coonintegralrep}
\end{align}
We note that the arguments that appear in the $q$-gamma functions of (\ref{coonq-gamma}) are the non-linear $q$-deformed Regge trajectories defined by
\begin{align}
\alpha_q(s) \equiv \frac{\ln{\sigma}}{\ln{q}} = \frac{\ln{(1-s (1-q))}}{\ln{q}}~,
\end{align}
in contrast to the linear Regge trajectories $s=\lim_{q \to 1} \alpha_q(s)$ that appear in the Veneziano amplitude (\ref{venezianoamp}). 

From the integral representation (\ref{coonintegralrep}), we can extract the residue polynomials on the massive poles at $s=[n]_q$
\begin{align}
\mathcal{A}^\textrm{C}(s,t) \rightarrow \frac{1}{s-[n]_q} \times R^C_{[n]_q}(t)~, 
\end{align}
where the $q$-integers $[n]_q$ are defined in (\ref{q-integers}).
Then unitarity is equivalent to the condition that each residue polynomial admits a partial wave expansion in terms of $D$-dimensional Gegenbauer polynomials 
\begin{align}
\label{eq:RBG}
R^C_{[n]_q}(t) = \sum_{j} B^{(D)}_{n,j}(q) \, G_j^{(D)}(t)~,
\end{align}
with all of the coefficient functions $B^{(D)}_{n,j}(q)$ non-negative.

To determine $R^C_{[n]_q}(t)$, we note that the singularities in $s$ all come from the region of integration near $z \to 0$. To that end, we Taylor expand the integrand around $z=0$ using $(1-qz)_q^{-\alpha_q(t)-1} = \sum_{k=0}^{\infty} a_k(q,t) z^k$, which implies that the Coon amplitude can be written as 
\begin{align}
\mathcal{A}^\textrm{C}(s,t) = \frac{q^{\alpha_q(s) \alpha_q(t) - \alpha_q(s) - \alpha_q(t)}}{[-\alpha_q(s)]_q} \sum_{k=0}^{\infty} a_k(q,t) \frac{1}{[k+1-\alpha_q(s)]_q}~,
\end{align}
where we have utilized the identity
\begin{align}
\int_{0}^{1} d_qz \,\, z^{-\alpha_q(s)+k} = \frac{1}{[k+1-\alpha_q(s)]_q}~. 
\end{align}
Evaluating the residue polynomial $R^C_{[n]_q}(t)$, we obtain
\begin{align}
R^C_{[n]_q}(t) &= \lim_{s \to [n]_q} (s - [n]_q) \mathcal{A}^\textrm{C}(s,t) = \frac{1}{[-n]_q}q^{\alpha_q(t)(n-1)-n}\lim_{s \to [n]_q} \sum_{k=0}^{\infty} a_k \frac{s-[n]_q}{[k+1-\alpha_q(s)]_q}~,
\label{residuefunc1}
\end{align}
where we have used the fact that $\alpha_q(s)=n$ in the limit $s \to [n]_q$. The only term that survives is the one corresponding to $k=n-1$, for which the residue polynomial reduces to
\begin{align}
 R^C_{[n]_q}(t) &= \frac{q^{\alpha_q(t)(n-1)-n}}{[-n]_q}\lim_{s \to [n]_q} a_{n-1} \frac{s-[n]_q}{[n-\alpha_q(s)]_q}~.
\end{align}
Manipulating the limit in the above expression and using the fact that $q^{\alpha_q(t)(n-1)-n} = \exp{ \alpha_q(t)(n-1) \, \log q - n \log q} = (1+(q-1)t)^{n-1}q^{-n}$, we obtain
\begin{align}
R^C_{[n]_q}(t) = a_{n-1} (q,t) \frac{q^n}{[n]_q} (1+(q-1)t)^{n-1}~,   
\end{align}
where we have used the property $[-n]_q = q^{-n}[n]_q$ of $q$-integers. We can now obtain the Taylor series coefficients $a_n$ as a contour integral around $z=0$ by Cauchy's theorem 
\begin{align}
a_n(q,t) = \oint_{z=0} \frac{dz}{2\pi i}  \frac{(1-qz)_q^{-\alpha_q(t)-1}}{z^{n+1}}~,    
\end{align}
which finally gives us the expression for the residue polynomial $R^C_{[n]_q}(t)$ as a contour integral around $z=0$:
\begin{align}
R^C_{[n]_q}(t) = (1+(q-1)t)^{n-1}\frac{q^n}{[n]_q} \oint_{z=0} \frac{dz}{2\pi i} \frac{(1-qz)_q^{-\alpha_q(t)-1}}{z^n}~. 
\label{residuepolynomial}
\end{align}

Next, we would like to compute the partial wave coefficients by expanding $R^C_{[n]_q}(t)$ in terms of the $D$-dimensional Gegenbauer polynomials\footnote{It might seem mathematically more natural to expand in the $q$-deformed Gegenbauer polynomials (also known as Rogers polynomials) \cite{gasper_rahman_2004}, but the physical interpretation of such an expansion would be unclear. Unitarity demands that the coefficient of each Gegenbauer polynomial must be non-negative; it is not clear what this fact implies for the coefficients of an expansion in $q$-deformed Gegenbauer polynomials. If a function admitting a non-negative expansion in terms of Gegenbauer polynomials is said to satisfy unitarity, then perhaps a function admitting a non-negative expansion in terms of $q$-deformed Gegenbauer polynomials should be said to satisfy $q$-nitarity.}  $G_j^{(D)}(x)$, where $x=\cos{\theta}$ and is related to $t$ as $t=\frac{s(x-1)}{2} = \frac{[n]_q (x-1)}{2}$ on the resonance $s=[n]_q$. We shall make use of the Rodrigues formula for $G_j^{(D)}(x)$:
\begin{align}
G_j^{(D)}(x) = (-1)^j \alpha_{j,D} (1-x^2)^{-\delta} \partial^j_x(1-x^2)^{\delta+j}~,   
\label{rodrigues}
\end{align}
where $\alpha_{j,D}= (2\delta+1)_j/(2^j j! (\delta+1)_j)$ with $\delta = \frac{D-4}{2}$. Now, orthogonality allows us to expand any smooth function $F(x)$ in terms of these Gegenbauer polynomials as $F(x)=\sum_{j=0}^{\infty} F_j G_j^{(D)}(x)$, where
\begin{align}
F_j = n_{j,D} \int_{-1}^1 dx (1-x^2)^{\delta} G_{j}^{(D)}(x) F(x)~,\quad n_{j,D} = j! \frac{2^{2 \delta} \left(j+\delta+\frac{1}{2}\right)\left[\Gamma\left(\delta+\frac{1}{2}\right)\right]^2}{\pi \Gamma(j+2\delta+1)}~.    
\end{align}
Using the Rodrigues formula (\ref{rodrigues}), one can integrate by parts $j$ times to obtain for the coefficients $F_j=\int_{-1}^1 dx (1-x^2)^{\delta+j} \partial_x^j F(x)$, where we have suppressed the overall positive factor $n_{j,D}$. We can apply this form to extract the coefficients in the Gegenbauer expansion of the residue polynomial which can be written as (\ref{eq:RBG}). Integrating by parts leaves us with the Gegenbauer coefficients $B^{(D)}_{n,j}(q)$ expressed as
\begin{equation}
    B^{D}_{n,j}(q) = c^{D}_{n,j}(q)\oint_{z=0} \frac{dz}{z^n}H_q(z)~,
    \label{Bqequation}
\end{equation}
where
\begin{equation}
    c^{D}_{n,j}(q) = q^n\frac{2^{D-3}}{[n]_q^{D+j-2}}\frac{(j+\frac{D-3}{2})\Gamma\left(\frac{D-3}{2}\right)}{\sqrt{\pi}\Gamma(j+\frac{D-2}{2})} > 0~, 
    \label{cqequation}
\end{equation}
is an overall, manifestly positive constant, and 
\begin{equation}
    H_q(z) = \int^{0}_{-[n]_q}dt (-t(t+[n]_q))^{\frac{D-4}{2}+j}\partial_t^j[(1-(1-q)t)^{n-1}(1-qz)_q^{-\alpha_q(t)-1}]\label{Hqequation}~.
\end{equation}
We note that in the limit $q \to 1$, the result (\ref{Hqequation}) is in perfect agreement with the $t$-integral given by (C.9) corresponding to the type-I superstring amplitude ($a=0$) in \cite{Arkani-Hamed:2022gsa}. Unfortunately, compared to the analysis in that paper, for general $q$ it appears difficult to further find representations that diagonalize the action of the differential operator $\partial_t^{j}$ on the $q$-function in (\ref{Hqequation}). This is a roadblock to further manipulating the $t$-integral to attain a double-contour representation along the lines of \cite{Arkani-Hamed:2022gsa}. Instead, we now highlight the utility of (\ref{Hqequation}) for recovering several existing results before moving on to discuss novel applications in section \ref{sec:reggetrajectories}.

\subsection{\texorpdfstring{The critical value $q=2/3$}{The critical value q=2/3}}

In this section we demonstrate that the coefficients $B_{n,j}^D(q)$ given in (\ref{Bqequation}) are non-negative in arbitrary dimension $D$ for $q \le 2/3$, recovering a result of \cite{Figueroa:2022onw} in the $\alpha_0\to 0$ limit.

Starting from (\ref{Bqequation}) and (\ref{Hqequation}) and applying the residue theorem we have
\begin{equation}
    B^{D}_{n,j}(q) = c^{D}_{n,j}(q)\frac{1}{[n-1]_q!}D_q^{(n-1)}H_q(0)~,
\end{equation}
where
\begin{equation*}
    \frac{1}{[n-1]_q!}D_q^{(n-1)}H_q(0) = \frac{(-1)^{n-1}q^{\frac{n(n-1)}{2}}}{[n-1]_q!}\int_{-[n]_q}^{0}dt (-t(t+[n]_q))^{\frac{D-4}{2}}\partial^j_t\prod_{k=1}^{n-1}\left(\frac{1-q^{-k}}{1-q}-t\right)~,
\end{equation*}
where $D_q^{(n-1)}$ denotes the Jackson $q$-derivative operated $(n-1)$-times with respect to $z$ and we have made use of the identity (\ref{derivativeex2}). Now making the substitution $t\to -[n]_qt$ and manipulating the finite product in the integrand yields the following result
\begin{align*}
     \frac{1}{[n-1]_q!}&D_q^{(n-1)}H_q(0) \\
     &=\frac{(-1)^{n-1+j}q^{\frac{n(n-1)}{2}}}{[n]_q!}[n]_q^{n+D+j-3}\int_{0}^{1}dt (t(1-t))^{\frac{D-4}{2}+j}\partial^j_t\prod_{k=1}^{n-1}\left(-\frac{q^{-k}-1}{1-q^n}+t\right)~,
\end{align*}
and therefore for the coefficients, one arrives at 
\begin{equation}
    B_{n,j}^{D}(q) = \frac{2^{D-3}(-1)^j}{\sqrt{\pi}}\frac{\left(\frac{D-3}{2}+j\right)\Gamma\left(\frac{D-3}{2}\right)}{\Gamma\left(\frac{D-2}{2}+j\right)}\frac{q^{\frac{n(n+1)}{2}}[n]_q^{n-1}}{[n]_q!}\int_{0}^{1}dt (t(1-t))^{\frac{D}{2}-2+j}\partial^j_t\prod_{k=1}^{n-1}\left(\frac{q^{-k}-1}{1-q^{n}}-t\right)~.
    \label{B-equation2}
\end{equation}
Finally we consider the substitution $t= \frac{1-x}{2}$, and the integral looks as follows\footnote{Up to a positive constant of proportionality that depends on $q,n$ and $D$.}
\begin{equation}
    B_{n,j}^{D}(q) \propto \int_{-1}^{1}dx (1-x^2)^{\frac{D}{2}-2+j}\partial^j_x\prod_{k=1}^{n-1}\left(\frac{2q^{-k}-3+q^n}{1-q^{n}}+x\right)~.
    \label{bnodprop}
\end{equation}
The above integral is manifestly positive if all roots of the polynomial in the product are negative. This is the case if $2q^{-k}-3+q^n \geq 0$ so if the minimum value of the LHS satisfies this then it will be positive for all $k$. So we take $k=1$ which implies $2q^{-1}-3+q^n \geq 0$, but since this is a decreasing function of $n$ it is minimised as $n \to \infty$ and we get 
\begin{align}
2q^{-1}-3 \geq 0 \implies q_{\text{crit}}=\frac{2}{3}~.  
\label{qcrit}
\end{align}
Therefore, we conclude that the Coon amplitude is manifestly unitary for $q \leq 2/3$ for any spacetime dimensions $D$, in agreement with the results of \cite{Figueroa:2022onw}.

\subsection{\texorpdfstring{Some coefficients at small $n$ and $j$}{Some coefficients at small n and j}}

Using the finite product formula (\ref{bnodprop}), we list in table \ref{tab:lowlyingcoeff} some of the low-lying partial wave coefficients for the Coon amplitude; for comparison, we list the corresponding coefficients for the Veneziano amplitude in table \ref{tab:lowlyingcoeffven}.
\begin{table}[ht]
    \centering
    \begin{tabular}{c|cccccc}
     &  $j=0$   &  $j=1$   & $j=2$    & $j=3$   \\ \hline
  $n=1$  & $q$ &   &   & & \\
  $n=2$   & $\frac{q^2(1-q)(2+q)}{2(1+q)}$ & $\frac{q^3}{2(D-3)}$  &   & & \\
  $n=3$   & $b^{D}_{3,0}(q)$ &$\frac{q^4(1-q)(1+3q+2q^2+q^3)}{2(D-3)(1+q)}$&$\frac{q^6(1+q+q^2)}{2(D-1)(D-3)(1+q)}$& & \\
  $n=4$ &$(1-q)b^{D}_{4,0}(q)$&$b^{D}_{4,1}(q)$&$(1-q)b^{D}_{4,2}(q)$&$\frac{3 q^{10}(1 + q)(1 + q^2)^2}{4 (D-3) (D^2-1) (1 + q + 
   q^2)}$&\\
\end{tabular}
    \caption{Values of $B^{D}_{n,j}(q)$ for Coon amplitude for low lying $(n,j)$. All the leading and sub-leading Regge coefficients are manifestly positive for $D\geq4$.  Expressions for the coefficients $b_{3,0}^{D}, b_{4,0}^{D},b_{4,1}^{D}$ and $b_{4,2}^{D}$ are too long to fit inside the table and are given in equations (\ref{eq:b30}-\ref{eq:b42}).}\label{tab:lowlyingcoeff}
    \end{table}

\begin{table}[ht]
    \centering
    \begin{tabular}{c|cccccc}
     &  $j=0$   &  $j=1$   & $j=2$    & $j=3$   \\ \hline
  $n=1$  & 1 &   &   & & \\
  $n=2$   & 0 & $\frac{1}{2(D-3)}$  &   & & \\
  $n=3$   & $\frac{10-D}{24(D-1)}$ &0&$\frac{3}{4(D-1)(D-3)}$& \\
  $n=4$ &$0$&$\frac{11-D}{12(D+1)(D-3)}$&$0$&$\frac{2}{(D^2-1)(D-3)}$&\\
\end{tabular}
    \caption{Values of $B^{D}_{n,j}$ for the Veneziano amplitude for low lying $(n,j)$. These results were listed for example in \cite{Arkani-Hamed:2022gsa}.}\label{tab:lowlyingcoeffven}
\end{table}
\noindent
In table \ref{tab:lowlyingcoeff} we have defined:
\begin{align}
b_{3,0}^D(q) &= \frac{q^3\left[4(D-1)+2q(D-1)-6q^2(D-1)-q^3(5D-6)-2q^4(D-2)+3Dq^5+2Dq^6+Dq^7\right]}{4(D-1)(1+q)(1+q+q^2)}~,\label{eq:b30}
\end{align}
\begin{align}
    b_{4,0}^D(q) &= \frac{q^4}{8(D-1) (1+q^2) (1+q+q^2)}\bigg[8(D-1) +4(D-1)q +4(D-1)q^2 \nonumber\\
    &+ 2(8 - 
      7D)q^3 + 2(10-7D)q^4
      +16(2-D)q^5 -3(3D-10)q^6+ 4(7+D)q^7\nonumber\\
      &+ (22+5D)q^8 + 8(2+ 
      D)q^9+5(2+D)q^{10} + 2(2+D)q^{11} + (2+D)q^{12}\bigg]~,\label{eq:b40}\\
      b_{4,1}^D(q) &= \frac{q^5}{8 (D-3) (1 + D) (1 + q+q^2) (1 + 
     q + q^2+q^3)}\bigg[4(1+D) + 12(1+D)q+12(1+D)q^2 \nonumber\\
     &-24(1+D)q^4- 3(6+7D)q^5- 2(4 +7D)q^6-3(D-2)q^7+ 12(2+D) q^8+ 9(2+D) q^9\nonumber\\ 
     & + 6(2+D)q^{10} + 3(2+D)q^{11}\bigg]~,\label{eq:b41}\\
     b_{4,2}^D(q) &= \frac{q^7 (1 + q^2) (2 + 6 q + 12 q^2 + 9 q^3 + 6 q^4 + 3 q^5)}{4 (D-3) (D-1)(1 + q + q^2)}~. \label{eq:b42}
\end{align}
The coefficients derived in the tables above are found to be in agreement with certain recent results in the literature:
\begin{itemize}
    \item The values of $B_{n,j}^D(q)$ listed in table \ref{tab:lowlyingcoeff} are in agreement with the results of \cite{Geiser:2022icl}, where the authors computed the closed-form expressions for the Coon amplitude partial-wave coefficients up to $n=3$. 
    \item An explicit formula for the coefficient $B_{n,j}^D(q)$, in fact for a four-parameter generalization of the Coon amplitude, was written as a nested four-fold sum in (67) of \cite{Cheung:2022mkw}.
    \item Moreover, it is easy to check that in the $q\to 1$ limit, table \ref{tab:lowlyingcoeff} reduces to the coefficients $B_{n,j}^D$ for the Veneziano amplitude listed in table \ref{tab:lowlyingcoeffven}. These were first derived in \cite{Arkani-Hamed:2022gsa} for arbitrary spacetime dimensions $D$ and in \cite{Maity:2021obe} for the limiting case of $D=4$. 
\end{itemize}
  Further, we notice manifest positivity of the leading and sub-leading Regge coefficients i.e., $B_{n,n-1}^D$ and $B_{n,n-2}^D$ for $n\in\{1,2,3,4\}$ in arbitrary $D$. We demonstrate the latter fact for any general $n$ in section \ref{sec:sregge} and then later around $q=1$ for any $j$ in the large $n$ limit in section \ref{sec:appliregge}.

\section{Regge Trajectories}
\label{sec:reggetrajectories}
In this section, we analyze the positivity of partial wave coefficients on Regge trajectories, where $j = n - k$ for some fixed $k$. The leading Regge trajectory has $j=n-1$, which is the highest spin for a given level $n$. We find that the leading and first sub-leading Regge coefficients are manifestly positive in all dimensions, but that unitarity breaks down in certain regions of $(q,D)$ parameter space starting with the sub-sub-leading Regge coefficients $B_{n,n-3}^D(q)$.

\subsection{Leading and sub-leading Regge coefficients}\label{sec:sregge}

We first determine the partial wave coefficients $B^{D}_{n,n-1}(q)$ of the Coon amplitude on the leading Regge trajectory. As before, our starting point is the contour expression for the partial wave coefficients (\ref{Bqequation}) with $(n,j)=(n,n-1)$: 
\begin{align}
B^{D}_{n,n-1}(q) = c^{D}_{n,n-1}(q) \frac{D_q^{(n-1)}H_q(0)}{[n-1]_q!}~.
\end{align}
Now, performing the Jackson $q$-derivative $(n-1)$-times on (\ref{Hqequation}) (with $j=n-1$) in the usual manner, we obtain for this derivative
\begin{align}
\frac{D_q^{(n-1)}H_q(0)}{[n-1]_q!} &= \frac{q^{\frac{n(n-1)}{2}}(n-1)!}{[n-1]_q!} \int_{-[n]_q}^0 dt (-t(t+[n]_q))^{\frac{D-6}{2}+n} \nonumber \\ 
&= \frac{q^{\frac{n(n-1)}{2}} (n-1)!}{[n]_q!} [n]_q^{D-4+2n} \frac{\Gamma\left(\frac{D}{2}-2+n\right)^2}{\Gamma(D-4+2n)}~,
\end{align}
where in reaching the final equality we have made a change of variable $t^{\prime} = -\frac{t}{[n]_q}$, and then evaluated the $t$-integral. Plugging in the value for the prefactor $c^D_{n,n-1}(q)$ leads us to the final expression for the leading Regge trajectory coefficient $B^D_{n,n-1}(q)$ which reads as
\begin{align}
B^D_{n,n-1}(q) &= 2^{2-2n} q^{\frac{n(n+1)}{2}} \frac{(n-1)![n]_q^{n-1}}{[n]_q!} \frac{\Gamma\left(\frac{D-3}{2}\right)}{\Gamma\left(\frac{D-5}{2}+n\right)}~ >0~.
\label{leadingreggetypeI}
\end{align}
We see that the above expression is clearly manifestly positive in arbitrary spacetime dimensions $D$. The above result is in agreement with the following special cases that have been recently derived in the literature:
\begin{itemize}
    \item The leading Regge coefficient for the Coon amplitude (\ref{leadingreggetypeI}) coincides with that of the Veneziano case as derived in \cite{Arkani-Hamed:2022gsa}. To see this, we take the $q \to 1$ limit of (\ref{leadingreggetypeI}) and see that this agrees with eq.~(4.38) in that reference with $\Delta=1$. 
    \item The formula above also coincides with the recently derived partial wave coefficients in \cite{Chakravarty:2022vrp} for the Coon amplitude in $D=4$. To make contact with their result, we set $n \to n+1$ for the relevant terms (as $j_{\textrm{max}}=n$ for the open, bosonic string which is the case considered in \cite{Chakravarty:2022vrp}) and take the $D \to 4$ limit of (\ref{leadingreggetypeI}). To that end, we obtain
\begin{align}
\lim_{D \to 4} B^D_{n,n}(q) = \frac{\sqrt{\pi}}{2^{2n}}  q^{\frac{n(n+1)}{2}} \frac{n!}{\Gamma\left(n+\frac{1}{2}\right)}\frac{[n]_q^{n-1}}{[n-1]_q!}~.
\label{reggeD=4}
\end{align}
Expressing the $q$-factorials as an infinite product and rearranging the terms in (\ref{reggeD=4}), we see that it exactly coincides with the expression in \cite{Chakravarty:2022vrp} with $\alpha_0=0$.
\item In the further special case of the Veneziano amplitude in $D=4$, the result (\ref{leadingreggetypeI}) in the $D=4, q \to 1$ limit coincides with the expression derived in \cite{Maity:2021obe}.  
\end{itemize}

We now move on to the sub-leading Regge trajectory where we can follow a similar prescription to determine the coefficient $B_{n,n-2}^D(q)$. Setting $j=n-2$ in (\ref{Bqequation}) and acting with the $q$-derivatives gives
\begin{equation}
    B_{n,n-2}^{D} = -c_{n,n-2}^{D}\frac{q^{\frac{n(n-1)}{2}} (n-2)!}{[n]_q!}\int_{-[n]_q}^{0}dt (-t(t+[n]_q))^{\frac{D}{2}+n-4} \sum_{k=1}^{n-1}\left(\frac{1-q^{-k}}{1-q}-t\right) ~.
\end{equation}
Considering a similar change of variables as before where $t^{\prime} = -\frac{t}{[n]_q}$ and computing the sum using $ \sum_{k=1}^{n-1}\left(\frac{q^{-k}-1}{1-q^n}-t\right) = \frac{q^{-n+1}}{1-q^n}-\frac{n}{1-q^n}-(n-1)t$ leaves us with the $t$-integral to evaluate. Evaluating this $t$-integral and inserting the expressions for the prefactor $c^{D}_{n,n-2}(q)$ from (\ref{cqequation}), one finally obtains for the sub-leading Regge trajectory coefficient $B_{n,n-1}^D(q)$ to be:
\begin{align}
\label{sub-leadingregge}
B_{n,n-2}^D(q) = 2^{3-2n} q^{\frac{n(n-1)}{2}} (1-q) \frac{[n]^{n-2}_q (n-2)!}{[n]_q!} \frac{\Gamma\left(\frac{D-3}{2}\right)}{\Gamma\left(\frac{D-7}{2}+n\right)} \times f(n,q)~,
\end{align}
where the polynomial $f(n,q)$ is defined as
\begin{align}
f(n,q)=\frac{2q+(n-1)(1-q)q^{2n}+ q^n(1-3n(1-q)-3q)}{(1-q)^3}~. 
\label{f(n,q)}
\end{align}
Although this form is clearly not manifestly positive, it can be written as
\begin{equation}
    f(n,q) = \sum_{k=1}^{n-2}(k+1)kq^k+\sum_{k=1}^{n}k(n-1)q^{2n-k-1} > 0~,
\end{equation}
which is manifestly positive, thereby proving that the sub-leading Regge coefficients $B^{D}_{n,n-2}$ are positive for \textit{all} $n, q$ and $D$.
We can easily see that the above expression (\ref{sub-leadingregge}) for the sub-leading Regge coefficient of the Coon amplitude approaches zero in the $q \to 1$ limit:
\begin{align}
\lim_{q \to 1} B_{n,n-2}^D = 0~,     
\end{align}
which is the expected result for the Veneziano amplitude for the case when $n+j \in 2\mathbb{Z}$ as shown in \cite{Arkani-Hamed:2022gsa,Maity:2021obe}.

\subsection{Sub-sub-leading Regge trajectory}\label{sec:ssregge}
We now move on to analyze the sub-sub-leading Regge coefficient $B^{D}_{n,n-3}(q)$. Starting again from (\ref{Bqequation}, \ref{Hqequation}) and following a similar prescription as in the previous sub-section, we have:
\begin{align}
    B_{n,n-3}^{D}(q) &= q^{\frac{n(n+1)}{2}} 2^{D-3}\frac{(n+\frac{D-9}{2})\Gamma\left(\frac{D-3}{2}\right)}{\sqrt{\pi}\Gamma(n+\frac{D-8}{2})}\frac{(n-3)!}{[n]_q!}[n]_q^{n-1}\nonumber\\
&~~~\times \int_{0}^{1}dt (t(1-t))^{\frac{D-10}{2}+n}\sum_{j=1}^{n-1}\sum_{k=j+1}^{n-1}\left(\frac{q^{-j}-1}{1-q^n}-t\right)\left(\frac{q^{-k}-1}{1-q^n}-t\right)~.
    \label{BofSSR}
\end{align}
Performing the double sum gives us the following rather complicated expression 
\begin{align}
B_{n,n-3}^D(q) &=
    2^{4-2n}q^{\frac{1}{2}n(n-3)}[n]_q^{n-1}\frac{(n-3)!}{[n]_q!}\frac{\Gamma\left(\frac{D-3}{2}\right)}{\Gamma\left(n+\frac{D-9}{2}\right)}\bigg[\frac{q^{2n}(n-2)(n-1)(D+2n-6)}{2(2n+D-7)}\nonumber\\
    &\hspace{5mm}+\frac{4q^{3}}{(1-q)^2(1+q)}+\frac{2n(n-1)q^{2n}}{(1-q^n)^2}+\frac{q^{2n}(4q-2n(2-n(1-q))(1+q))}{(1-q^2)(1-q^n)}\nonumber\\
    &\hspace{5mm}-\frac{2q^{1+n}(3n(1+q)-2(2+q))}{1-q^2}
    \bigg]\label{ssrB}~.
\end{align}
One can easily verify that in taking the $q\to1$ limit, one arrives at the following result 
\begin{equation}
    B_{n,n-3}^D(q)\overset{q\to1}{\longrightarrow} \frac{2^{5 - 2 n} n^{n-3}(7 - D + n) (D-9 + 2 n) \Gamma\left(
  \frac{D-3}{2}\right)}{48\,\Gamma\left(\frac{D-5}{2}+n\right)}~,
\end{equation}
and then using the Legendre duplication formula:
\begin{equation}
    \Gamma\left(\frac{z}{2}\right)\Gamma\left(\frac{z+1}{2}\right) = 2^{1-z}\sqrt{\pi}\Gamma\left(z\right)~,\qquad \forall z\in \mathbb{C}/\mathbb{Z}_{\leq0}
\end{equation}
one obtains the result derived in \cite{Arkani-Hamed:2022gsa}
for the type-I superstring amplitude for $\Delta = 3$. 

\begin{figure}[ht]
    \centering
    \qquad
     \subfloat[\centering The coefficient $B^D_{5,2}(q)$]{{\includegraphics[scale=0.45]{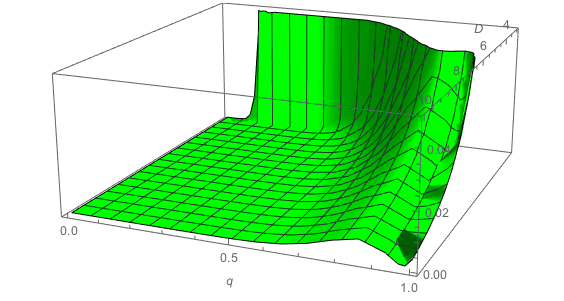}}}%
    \subfloat[\centering Non-unitary regions for $B^D_{n,n-3}(q)$]{{\includegraphics[scale=1.35]{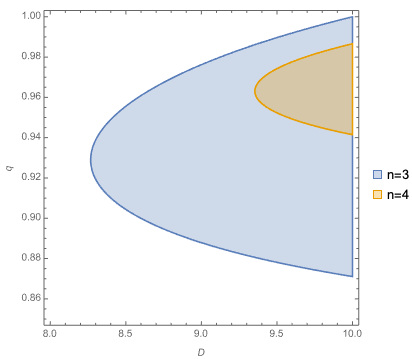}}}%
    \caption{On moving to the next sub-sub-leading Regge coefficient $B^D_{5,2}(q)$ in the plot (a), we see that the negative patch disappears and the surface becomes positive for all values of $q \in [0,1]$ and $D \in [4,10]$. The region plot (b) shows more clearly the negative patches for $B^D_{n,n-3}(q)$ in $(q,D)$ parameter space. The tip of the non-unitary region lies at $q = 0.9289\ldots$, $D=8.2672\ldots$; the exact numbers are roots of sixth order polynomial equations following from (\ref{ssrB}) with $n=3$.}
    \label{regplot}
\end{figure}

Like the sub-leading Regge coefficients, the sub-sub-leading case (\ref{ssrB}) is not manifestly positive for general $q$. Moreover, on plotting the first few low-level coefficients (see figure \ref{B3,0plot3D}) we see that for certain values of $q$ and $D$, the coefficient for the $n=3$ level starts becoming negative (indicated in the figure by the red patch). We note from figure \ref{B3,0plot3D} that the coefficient $B^D_{3,0}(q)$ vanishes as $q \to 1,D \to 10$ which is the behavior one would expect for the Veneziano amplitude coefficient $B^D_{3,0}=\frac{10-D}{24(D-1)}$ (from table \ref{tab:lowlyingcoeffven}). Similarly for the $n=4$ coefficient, we continue to see a small red patch around $q=1$ and $D=10$ (albeit smaller), but the patch completely disappears for the $n=5$ coefficient (see figure \ref{regplot}). 
\begin{figure}[ht]
    \centering
     \subfloat[\centering The coefficient $B^D_{3,0}(q)$]{{\includegraphics[scale=0.40]{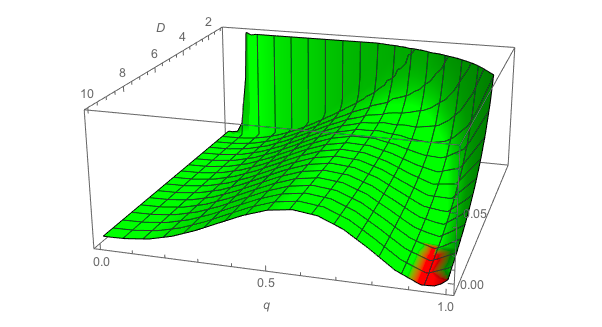}}} %
     \subfloat[\centering The coefficient $B^D_{4,1}(q)$]{{\includegraphics[scale=0.40]{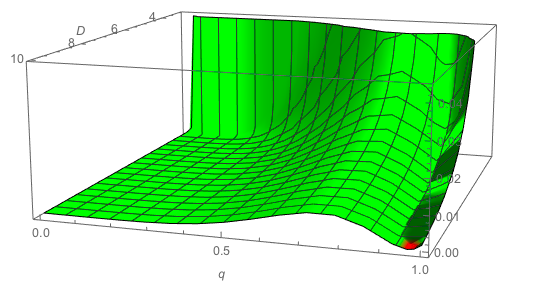}}} %
    \caption{The green region indicates positivity while the red region shows where the indicated partial wave coefficients become negative. We also note from the plot that the coefficient $B^D_{3,0} \to 0$ as $q \to 1, D \to 10$ which is consistent with the expression $B^D_{3,0}$ obtained in \cite{Arkani-Hamed:2022gsa}. On the right, we continue to see the negative patch (although smaller) in the plot (b) very close to the critical dimension $D=10$ of the Veneziano amplitude ($q \to 1$).} 
    \label{B3,0plot3D}
\end{figure}
We will show in section \ref{sec: smallqdeform} that all Regge coefficients of the Coon amplitude for large values of $n$ are manifestly positive in a neighborhood of $q=1$. 
%\newpage
\subsection{Leading versus sub-leading Regge behaviour as \texorpdfstring{$n \to \infty$}{n to infinity}}\label{sec:lvsssr}
We now compare the leading, sub-leading, and sub-sub-leading Regge coefficients that we calculated in the last two sections for various values of level $n$. 
\begin{figure}[ht]
    \centering
    \subfloat[\centering]{{\includegraphics[scale=0.35]{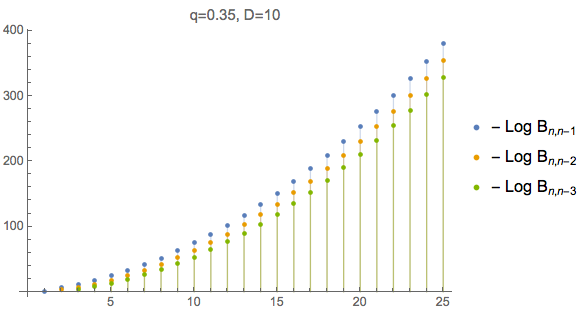}}}%
    \qquad
    \subfloat[\centering]{{\includegraphics[scale=0.35]{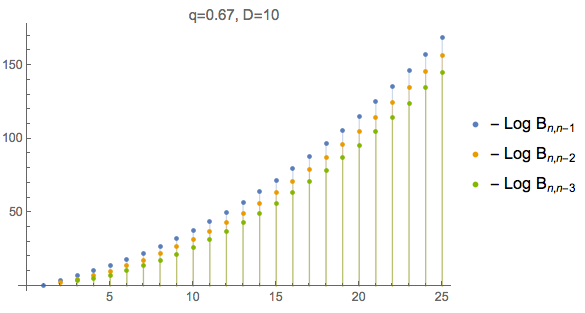}}}%
    \qquad
    \subfloat[\centering]{{\includegraphics[scale=0.35]{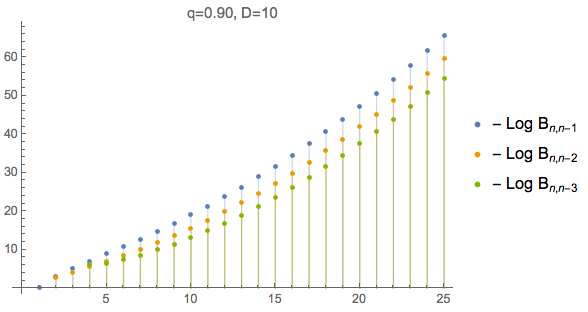}}}%
    \qquad
     \subfloat[\centering]{{\includegraphics[scale=0.35]{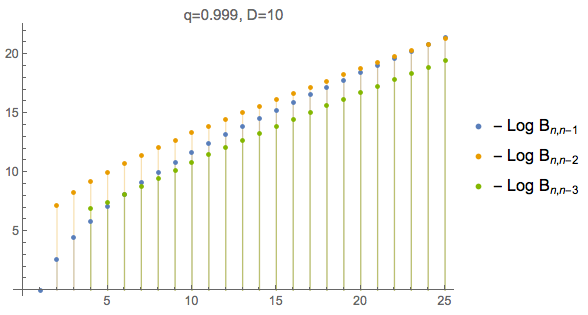}}}%
    \caption{Coefficients of leading, sub-leading and sub-sub-leading Regge trajectories for various values of $q$ for $\alpha_0=0$. From plot (a), (b) and (c) it is clear that $B^{10}_{n,n-3}>B^{10}_{n,n-2}>B^{10}_{n,n-1}$ $\forall n$. However, for plot (d) this is only true for larger values of $n$.}
    \label{fig:leadingvssr}
\end{figure}
In figure \ref{fig:leadingvssr} we compare these coefficients for the first 25 levels for various values of $q$ in $D=10$. We can see that unless $q$ is very close to 1, the sub-leading coefficients always dominate over the leading ones. This is clearly the case with the first three sub-plots in figure \ref{fig:leadingvssr}. Therefore, away from the Veneziano case, our numerical results support the hypothesis that $B^{D}_{n,n-\Delta_1}>B^{D}_{n,n-\Delta_2}$ for $\Delta_1>\Delta_2$ as $n \to \infty$. Thus the positivity of the leading Regge trajectory is sufficient to conclude that all Regge trajectories are positive-definite in the asymptotic limit. In fact, one can easily verify numerically that this true $\forall$ $D\geq4$. However, as one reaches closer to the Veneziano case (the last subplot in the figure) this pattern breaks down and one needs to go to larger and larger values of $n$ to recover this trend. In such a case, the pattern depends greatly on the order of limits and there may be some doubt about whether positivity indeed holds on all Regge trajectories as $n \to \infty$. In order to close this gap, we turn our attention in the next section to an analysis of the $q \approx 1$ region of parameter space, showing that the Regge coefficients are indeed positive for any $D$ as $n \to \infty$.

\section{Coon amplitude around \texorpdfstring{$q=1$}{q=1}}
\label{sec: smallqdeform}
So far, our single-contour expression (\ref{Bqequation}) for the partial wave coefficients $B^D_{n,j}(q)$ of the Coon amplitude has found applications in evaluating certain low-lying coefficients and obtaining analytic, manifestly positive expressions for the leading and sub-leading Regge trajectories. It has also helped us obtain a closed-form expression for the sub-sub-leading Regge trajectory which showed us the first signs of the breakdown of positivity. To gain further understanding of the Gegenbauer coefficients, we must strive towards obtaining a double-contour integral representation for them, in a spirit close to the analysis in \cite{Arkani-Hamed:2022gsa} for the Veneziano amplitude. In moving towards a double-contour integral representation for the Coon amplitude coefficients, we encounter difficulties mainly arising from technical obscurities in dealing with $q$-deformed quantities in the integrand. While a double-contour integral representation eludes us, we attempt to consider small deviations around the Veneziano amplitude by Taylor expanding our single-contour expression (\ref{Bqequation}) around $q=1$, up to the first order. As we shall see in this section, this would help us obtain a $\mathcal{O}(1-q)$ correction to the double-contour integral representation of the Veneziano coefficients \cite{Arkani-Hamed:2022gsa}. The upshot is that we get a crude understanding of the structures one would naively expect in a generic double-contour integral representation for the Coon amplitude (in a higher-order expansion). Our $\mathcal{O}(1-q)$ double-contour expression will also help us understand applications related to the asymptotics of these coefficients which we show to be manifestly positive, thus validating our discussion at the end of section \ref{sec:lvsssr} on analytic grounds. 

\subsection{Towards a double-contour representation}
To help us obtain a $\mathcal{O}(1-q)$ double-contour correction to the Veneziano amplitude ($q=1$), we start by considering small deviations around $q=1$. To that end, we perform a Taylor expansion of our single-contour expression (\ref{Bqequation}) around $q=1$ up to the first order. We refer the reader to Appendix \ref{sec:taylorexpappen} for the Taylor expansion of the various $q$-deformed quantities that appear in the integrand of (\ref{Bqequation}). We achieve this by dividing the integrand into two parts: i) the first order expansion of the portion inside the derivative of (\ref{Hqequation}) is shown explicitly in (\ref{integrandpart2}), ii) while the expansion of the remaining part of the integrand along with measure has been dealt with in (\ref{integrandpart1}). We can try to condense these integrals by introducing the $H$-integral along the lines of \cite{Arkani-Hamed:2022gsa} as follows\footnote{We note that our definition of the $H$-integral slightly differs from \cite{Arkani-Hamed:2022gsa} by an extra factor of $(1-z)$ in the denominator and the inclusion of the integer parameters $r,s$.}:
\begin{align}
H_{n,j}^{(r,s)}(t^{\prime}) = \int^{t^\prime} dt \frac{(-t)^{\frac{D-4}{2}+j+r} (t+n)^{\frac{D-4}{2}+j+s}}{(1-z)^{t+1}}~, \quad r,s \in \mathbb{Z}~.   
\label{h-integralmaintext}
\end{align}
We leave the exact details of manipulating these $H$-integrals (where one can express its integrand as a total derivative) to Appendix \ref{sec:taylorexpappen}. However, to perform an expansion of the Gegenbauer coefficients (\ref{Bqequation}), we must find an appropriate contour expression around $z=0$. To that end, we define a generalized basis of integrals which we term as the $\mathcal{H}$-integrals:
\begin{equation}
    \mathcal{H}^{(a,b,c,r,s)}_{n,j}(t') := \oint \frac{dz}{2\pi i}\frac{1}{z^{n+a}}\frac{1}{(1-z)^b}(-\log(1-z))^{j+c}H^{(r,s)}_{n,j}(t')~,
    \label{H-basistext}
\end{equation}
where $a \in \mathbb{Z}_{\leq0}$ and ${b,c,r,s} \in \mathbb{Z}$ parameterize the above set of integrals. We observe that these integrals satisfy the following property 
\begin{align}
    \mathcal{H}^{(a,b,c,r,s)}_{n,j}(-n) &= (-1)^{n+j+a+c}\oint \frac{dz}{z^{n+a}}(1-z)^{a+b}(-\log(1-z))^{j+c}H^{(s,r)}_{n,j}(0)\nonumber\\
    &= (-1)^{n+j+a+c} \mathcal{H}^{(a,-a-b,c,s,r)}_{n,j}(0)~,
    \label{hbasisidtext}
\end{align}
with the details of the derivation presented in Appendix \ref{sec:taylorexpappen}. Making use of the above identity and performing certain algebraic manipulations (for the explicit details we refer the reader again to Appendix \ref{sec:taylorexpappen}), we finally obtain a double-contour integral representation for the $\mathcal{H}$-integral expressed as\footnote{In \cite{Arkani-Hamed:2022gsa}, the coefficient $B^{\textrm{open}}_{n,j}$ is proportional to the integral $\mathcal{H}^{(0,0,0,0,0)}_{n,j}(0)$ which exactly matches with their result (C.17).}
\begin{align}
\mathcal{H}^{(a,b,c,r,s)}_{n,j}(0) &= (-1)^{r+s} \Gamma\left(j+r+\frac{D-2}{2}\right) \Gamma\left(j+s+\frac{D-2}{2}\right) \nonumber \\
&~~~~~~\times \oint_{u=0} \frac{du}{2 \pi i} \oint_{v=0} \frac{dv}{2 \pi i} \frac{(v-u)^{j+c} e^{-bu} e^{(a+b)v}}{(uv)^{\frac{D-2}{2}+j} u^r v^s (e^v-e^u)^{n+a}}~. 
\label{double-contourmasterintegraltext}
\end{align}
We now wish to express our first-order expansions of the two terms in the integrand, namely (\ref{integrandpart1}) and (\ref{integrandpart2}), in terms of the above-introduced basis integrals $\mathcal{H}^{(a,b,c,r,s)}_{n,j}(0)$. The contour integral of (\ref{integrandpart1}) is rewritten in the above basis as 
\begin{align}
\oint_{z=0} \frac{dz}{2 \pi i} \frac{1}{z^n} (\ref{integrandpart1}) &=
    (1-(-1)^{n+j})\mathcal{H}^{(0,0,0,0,0)}_{n,j}\nonumber\\ &-(1-q)\bigg\{\frac{n-1}{2}\left(\frac{D-2}{2}+j\right)(1-(-1)^{n+j})\mathcal{H}^{(0,0,0,0,0)}_{n,j} \nonumber \\
    &+\frac{n-1}{2}\left[\mathcal{H}^{(0,0,1,1,0)}_{n,j}+(-1)^{n+j}\mathcal{H}^{(0,0,1,0,1)}_{n,j}\right]\nonumber\\
    &+\left(\frac{D-4}{2}+j\right)\frac{n(n-1)}{2}\left[\mathcal{H}^{(0,0,0,0,-1)}_{n,j}-(-1)^{n+j}\mathcal{H}^{(0,0,0,-1,0)}_{n,j}\right]\nonumber\\
    &+\frac{n-1}{2}\left(\frac{D-4}{2}+j\right)\left[\mathcal{H}^{(0,0,0,1,-1)}_{n,j}-(-1)^{n+j}\mathcal{H}^{(0,0,0,1,-1)}_{n,j}\right]\bigg\}~, 
    \label{term1}
\end{align}
where we have made use of the identity (\ref{hbasisidtext}) to rewrite $\mathcal{H}_{n,j}^{(0,0,0,0,0)}(-n)$ in terms of the basis integrals.
Similarly, we have for the contour integral of (\ref{integrandpart2}) around $z=0$:
\begin{align}
\oint_{z=0} \frac{dz}{2 \pi i} \frac{1}{z^n} (\ref{integrandpart2}) &=  (1-(-1)^{n+j})\mathcal{H}^{(0,0,0,0,0)}_{n,j} -(1-q) \bigg\{\left(nj-\frac{j(j+1)}{2}\right) \mathcal{H}^{(0,0,-1,0,0)}_{n,j} (1+(-1)^{n+j}) \nonumber \\
&-\frac{1}{2} \left[\mathcal{H}^{(-1,1,0,2,0)}_{n,j}+(-1)^{n+j}\mathcal{H}^{(-1,0,0,0,2)}_{n,j}+\mathcal{H}^{(0,0,1,2,0)}_{n,j}+(-1)^{n+j}\mathcal{H}^{(0,0,1,0,2)}_{n,j} \right] \nonumber \\
&+j \bigg[\mathcal{H}^{(-1,1,-1,1,0)}_{n,j}-(-1)^{n+j}\mathcal{H}^{(-1,0,-1,0,1)}_{n,j}+\mathcal{H}^{(0,0,0,1,0)}_{n,j}-(-1)^{n+j}\mathcal{H}^{(0,0,0,0,1)}_{n,j}\bigg]\nonumber\\
&-\frac{j(j-1)}{2} \left[\mathcal{H}^{(-1,1,-2,0,0)}_{n,j}+(-1)^{n+j}\mathcal{H}^{(-1,0,-2,0,0)}_{n,j}\right]\bigg\}~,
\label{term2}
\end{align}
where we note that all the master $\mathcal{H}$-integrals in the above two equations are evaluated at $z=0$, i.e., $\mathcal{H}^{(a,b,c,r,s)}_{n,j}(0)$. We see that the first terms in both (\ref{term1}) and (\ref{term2}) coincide with that obtained for the Veneziano amplitude in \cite{Arkani-Hamed:2022gsa}. The remaining terms arise from considering $\mathcal{O}(1-q)$ corrections. 

Now, an expression that shows up quite often in relations (\ref{term1}) and (\ref{term2}) is of the form $\mathcal{H}^{(a,b,c,r,s)}_{n,j}-(-1)^{n+j+a+c}\mathcal{H}^{(a,-a-b,c,s,r)}_{n,j}$, which we can further simplify by combining the two terms into one expression. First, we consider an interchange in $u\leftrightarrow v$, $r\leftrightarrow s$ and $b\to -a-b$ in (\ref{double-contourmasterintegraltext}), by virtue of which we arrive at the following
\begin{align}
     \mathcal{H}^{(a,-a-b,c,s,r)}_{n,j}(0) &= (-1)^{r+s+n+j+a+c} \Gamma\left(j+r+\frac{D-2}{2}\right) \Gamma\left(j+s+\frac{D-2}{2}\right) \nonumber \\
&~~~~~~\times \oint_{v=0} \frac{dv}{2 \pi i} \oint_{u=0} \frac{du}{2 \pi i} \frac{(v-u)^{j+c} e^{-bu} e^{(a+b)v}}{(uv)^{\frac{D-2}{2}+j} u^r v^s (e^v-e^u)^{n+a}}~.
\label{identity1}
\end{align}
But, we then notice that
\begin{align}
    \mathcal{H}^{(a,b,c,r,s)}_{n,j}-&(-1)^{n+j+a+c}\mathcal{H}^{(a,-a-b,c,s,r)}_{n,j}\nonumber\\
    = &(-1)^{r+s} \Gamma\left(j+r+\frac{D-2}{2}\right) \Gamma\left(j+s+\frac{D-2}{2}\right)\nonumber\\
    &~~\times \left(\oint_{u=0}\frac{du}{2 \pi i}\oint_{v=0}\frac{dv}{2 \pi i}-\oint_{v=0}\frac{dv}{2 \pi i}\oint_{u=0}\frac{du}{2 \pi i}\right)\frac{(v-u)^{j+c} e^{-bu} e^{(a+b)v}}{(uv)^{\frac{D-2}{2}+j} u^r v^s (e^v-e^u)^{n+a}}\nonumber\\
    = &(-1)^{r+s+1} \Gamma\left(j+r+\frac{D-2}{2}\right) \Gamma\left(j+s+\frac{D-2}{2}\right)\nonumber\\
    &~~\times \oint_{u=0}\frac{du}{2 \pi i}\oint_{v=u}\frac{dv}{2 \pi i}\frac{(v-u)^{j+c} e^{-bu} e^{(a+b)v}}{(uv)^{\frac{D-2}{2}+j} u^r v^s (e^v-e^u)^{n+a}}~, \label{identity2}
\end{align}
where we made use of the identity familiar to us from two-dimensional CFT for contour integrals (for example, refer to section 6.1.2 of \cite{DiFrancesco:1997nk})
\begin{equation}
    \oint_{v=0}\frac{dv}{2 \pi i}\oint_{u=0}\frac{du}{2 \pi i}-\oint_{u=0}\frac{du}{2 \pi i}\oint_{v=0}\frac{dv}{2 \pi i} = \oint_{u=0}\frac{du}{2 \pi i}\oint_{v=u}\frac{dv}{2 \pi i}~. \label{commresidue}
\end{equation}
Observing the above identities, we immediately see that the number of terms in (\ref{term1}, \ref{term2}) can be exactly reduced to half if we make the following choice for our basis integrals:
\begin{align}
\mathcal{I}^{(a,b,c,r,s)}_{n,j}(0) &= (-1)^{r+s+1} \Gamma\left(j+r+\frac{D-2}{2}\right) \Gamma\left(j+s+\frac{D-2}{2}\right) \nonumber \\
&~~~~~~\times \oint_{u=0} \frac{du}{2 \pi i} \oint_{v=u} \frac{dv}{2 \pi i} \frac{(v-u)^{j+c} e^{-bu} e^{(a+b)v}}{(uv)^{\frac{D-2}{2}+j} u^r v^s (e^v-e^u)^{n+a}}~. \label{double-contourmasterintegral2}
\end{align}
we therefore classify the $\mathcal{I}$-integrals (\ref{double-contourmasterintegral2}) as our new basis integrals. 
Finally rewriting (\ref{term1}) and (\ref{term2}) in terms of our new basis of $\mathcal{I}$-integrals (\ref{double-contourmasterintegral2}) and combining it with the Taylor expansion of the coefficients, we arrive at the following more compact expression
\begin{align}
    B^{D}_{n,j}(q) &= c^{D}_{n,j}\bigg\{\mathcal{I}^{(0,0,0,0,0)}_{n,j}+(1-q)\bigg[\left(\frac{D-2}{4}(n-1)-n\right)\mathcal{I}^{(0,0,0,0,0)}_{n,j}\nonumber\\
    &-\frac{n-1}{2}\mathcal{I}^{(0,0,1,1,0)}_{n,j}-\left(\frac{D-4}{2}+j\right)\frac{n(n-1)}{2}\mathcal{I}^{(0,0,0,0,-1)}_{n,j}\nonumber\\
    &-\frac{n-1}{2}\left(\frac{D-4}{2}+j\right)\mathcal{I}^{(0,0,0,1,-1)}_{n,j}-\left(nj-\frac{j(j+1)}{2}\right) \mathcal{I}^{(0,0,-1,0,0)}_{n,j}\nonumber\\
    &+\frac{1}{2} \left(\mathcal{I}^{(-1,1,0,2,0)}_{n,j}+\mathcal{I}^{(0,0,1,2,0)}_{n,j}\right)-j \left(\mathcal{I}^{(-1,1,-1,1,0)}_{n,j}+\mathcal{I}^{(0,0,0,1,0)}_{n,j}\right)\nonumber\\
    &+\frac{j(j-1)}{2} \mathcal{I}^{(-1,1,-2,0,0)}_{n,j}\bigg]\bigg\}~. \label{BintermsofI}
\end{align}
Rewriting the above explicitly in terms of the double contour integral representation (\ref{double-contourmasterintegral2}), we get
\begin{align}
    B^{D}_{n,j}(q) &= -c^{D}_{n,j}(0)\Gamma\left(j+\frac{D-2}{2}\right)^2\oint_{u=0} \frac{du}{2 \pi i} \oint_{v=u} \frac{dv}{2 \pi i} \frac{(v-u)^{j-2}}{(uv)^{\frac{D-2}{2}+j}(e^v-e^u)^{n}}\mathscr{F}^{D}_{n,j}(u,v;q)~,
    \label{firstordercorrecB}
\end{align}
where the function $\mathscr{F}^{D}_{n,j}(u,v;q)$ is given by the relation
\begin{align}
\mathscr{F}^{D}_{n,j}(u,v;q) &= (v-u)^2 +(1-q) \bigg[\left(\left(\frac{D-2}{4}\right)(n-1)-n\right)(v-u)^2\nonumber\\
&+ \frac{n-1}{2}\bigg(j+\frac{D-2}{2}\bigg) \frac{(v-u)^3}{u} + \left(j+\frac{D-4}{2}\right) \frac{n(n-1)}{2} v \nonumber \\
&- \frac{n-1}{2} \frac{(v-u)^2 v}{u} - \left(nj-\frac{j(j+1)}{2}\right) (v-u) \nonumber \\
&+\frac{1}{2u^2}\left(j+\frac{D}{2}\right)\left(j+\frac{D-2}{2}\right)\left[(v-u)^3+(v-u)^2(e^{v-u}-1)\right] \nonumber \\
&+\frac{1}{u}j\left(j+\frac{D-2}{2}\right)\left[(e^{v-u}-1)(v-u)+1\right]+\frac{j(j-1)}{2}(e^{v-u}-1)\bigg]+\mathcal{O}(1-q)^2~, \label{fpolynomial}
\end{align}
where as before, the first term in (\ref{fpolynomial}) coincides with that obtained for the Veneziano amplitude in \cite{Arkani-Hamed:2022gsa} while the remaining terms arise from considering $\mathcal{O}(1-q)$ corrections.

\subsection{Applications}
Equipped with the first order $(1-q)$ correction to the double-contour integral representation for the Veneziano amplitude \cite{Arkani-Hamed:2022gsa}, we see glimpses of the structures one can expect as one flows away from $q=1$ from the relations (\ref{firstordercorrecB}, \ref{fpolynomial}). While this double-contour correction has little use in studying manifest unitarity around $q=1$\footnote{In fact we observe that studying positivity on a term-by-term basis leads us to conclude that the first order correction is not manifestly positive in any dimension $D$. This only concludes that not all terms in the Taylor expansion of $\mathscr{F}^D_{n,j}$ are manifestly positive, but that does not rule out the possibility for the function itself to be positive for all $q\in [0,1]$ once the entire series is summed over.}, we put this relation to use by studying the asymptotics of the Gegenbauer coefficients in two regimes: i) for fixed spin $j$, and ii) for fixed $\Delta = n-j$ (Regge trajectories).

\subsubsection{Fixed \texorpdfstring{$j$}{j} asymptotics} \label{sec:fixedjasymp}
We use our first order $q$-extension to the double contour integral representation for studying asymptotics (i.e., the large $n$ limit) at fixed spin $j$. We start by writing the $\mathcal{H}$-integrals 
\begin{align}
\mathcal{H}^{(a,b,c,r,s)}_{n,j}(0) \propto \oint_{u=0}\frac{du}{2 \pi i}\oint_{v=0}\frac{dv}{2 \pi i}\frac{(v-u)^{j+c} e^{-bu} e^{(a+b)v}}{(uv)^{\frac{D-2}{2}+j} u^r v^s (e^v-e^u)^{n+a}}~,  
\label{hintegralprop}
\end{align}
and simply change variables to 
\begin{align}
u=\log(1-x)~, \qquad v=\log(1-y)~,
\end{align}
to obtain the integral
\begin{align}
\mathcal{H}^{(a,b,c,r,s)}_{n,j}(0) \propto \oint_{x=0}\frac{dx}{2 \pi i}\oint_{y=0}\frac{dy}{2 \pi i}& \frac{1}{(1-x)^{1+b} (1-y)^{1-a-b} \log(1-x)^{\frac{D-2}{2}-c+r} \log(1-y)^{\frac{D-2}{2}-c+s}} \nonumber \\
&~~~~~\times \frac{1}{(x-y)^{n-j-c+a}} \left(\frac{\frac{1}{\log(1-x)}-\frac{1}{\log(1-y)}}{x-y}\right)^{j+c}~.    
\end{align}
Each of the three factors in the above integrand is a function with strictly positive Taylor coefficients when expanded first around $y=0$ followed by $x=0$. Such a function is termed as a \textit{positive} function. As a result, the coefficients picked out by the residue integral will be positive as well. We manipulate each of the factors in the above integrand one-by-one. The positivity of the first factor is positive based on the arguments presented in Appendix A of \cite{Arkani-Hamed:2022gsa}. Specifically, in the factor
\begin{align}
\log(1-y)^{1-\frac{D}{2}+c-s} = (-y)^{1-\frac{D}{2}+c-s} + \sum_{m=0}^{\infty} c_m y^m \approx (-y)^{1-\frac{D}{2}+c-s}~,
\label{factor1}
\end{align}
we note that only the leading term contributes in the $n \to \infty$ limit. For the binomial expansion\footnote{We note that the binomial coefficient in (\ref{binom1}) is always positive for the $\mathcal{H}$-integrals contributing at $\mathcal{O}(1-q)$. At subsequent higher orders, this term would have to be dealt with differently.} 
\begin{align}
(x-y)^{-n+j+c-a} &= \sum_{m=0}^{\infty} \binom{n+a-j-c+m-1}{m} x^{-n-a+j+c-m} y^m \nonumber \\
&\approx \binom{2n+2a-2j-2c-2}{n+a-j-c-1} x^{-2n-2a+2j+2c+1} y^{n+a-j-c-1}~, 
\label{binom1} 
\end{align}
where we have used the fact that the maximum binomial coefficient $\binom{n}{m}$ is obtained for $m=\lfloor{\frac{n}{2}}\rfloor$ and contributes solely to the sum in the large $n$ limit. The third factor in the integrand admits an expansion whose coefficients are all positive: 
\begin{align}
\left(\frac{\frac{1}{\log(1-x)}-\frac{1}{\log(1-y)}}{x-y}\right)^{j+c} = \frac{1}{(xy)^{j+c}} + \sum_{m=0}^{\infty} c_m \sum_{k=0}^{m-1} x^{m-1-k} y^k \approx \frac{1}{(xy)^{j+c}}~,
\label{factor3}
\end{align}
where only the first term survives in the asymptotic regime. Using the above results and performing the $y$-integral where we pick up the residue at $y=0$\footnote{When computing the residue in $y$, the greatest contribution comes from the highest possible power from (\ref{binom1}) as the binomial coefficients grow rapidly with $n$. For the other two factors (\ref{factor1}, \ref{factor3}), we simply pick up their leading coefficients in the $n \to \infty$ limit.}, we get
\begin{align}
\mathcal{H}^{(a,b,c,r,s)}_{n,j} &\sim \frac{n^{j+s+\frac{D-4}{2}}}{\left(j+s+\frac{D-4}{2}\right)!} \oint_{x=0}\frac{dx}{2 \pi i} \frac{x^{-n-a-j-s-\frac{D-4}{2}}}{(1-x)^{1+b} (-\log(1-x))^{\frac{D-2}{2}+r-c}}~.
\end{align}
where we have used Stirling's approximation on the factorials $n! \sim \sqrt{2 \pi n} \left(\frac{n}{e}\right)^n$. 
We first deform the above contour that runs around $0$ to the Hankel contour $\mathcal{C}$ that runs from below around the branch cut (see Figure VI.2 in \cite{DBLP:books/daglib/0023751} for exact details) and then perform a change of variables $x = 1+\frac{t}{n}$ to obtain
\begin{align}
\mathcal{H}^{(a,b,c,r,s)}_{n,j} &\sim \frac{n^{j+s+\frac{D-4}{2}}}{\left(j+s+\frac{D-4}{2}\right)!} \oint_{t=0}\frac{dt}{2 \pi i} \frac{1}{n} \frac{\left(1+\frac{t}{n}\right)^{-n-a-j-s-\frac{D-4}{2}}}{\left(-\frac{t}{n}\right)^{1+b} \left(-\log\left(-\frac{t}{n}\right)\right)^{\frac{D-2}{2}+r-c}} \nonumber \\
&= \frac{(-1)^{1+b}n^{j+s+\frac{D-4}{2}+b}}{\left(j+s+\frac{D-4}{2}\right)! (\log n)^{\frac{D-2}{2}+r-c}} \oint_{t=0}\frac{dt}{2 \pi i} \frac{\left(1+\frac{t}{n}\right)^{-n-a-j-s-\frac{D-4}{2}}}{t^{1+b} \left(1-\frac{\log(-t)}{\log(n)}\right)^{\frac{D-2}{2}+r-c}}~.  
\end{align}
The above integrand converges uniformly for any bounded domain on the $t$-plane and hence we replace it with its value at large $n$ which leads us to
\begin{align}
\mathcal{H}^{(a,b,c,r,s)}_{n,j} &\sim \frac{(-1)^{1+b}n^{j+s+\frac{D-4}{2}+b}}{\left(j+s+\frac{D-4}{2}\right)! (\log n)^{\frac{D-2}{2}+r-c}} \oint_{t=0}\frac{dt}{2 \pi i} \frac{e^{-t}}{t^{1+b}}~.   
\end{align}
Performing the final contour integral and restoring the proportionality factors we omitted in writing (\ref{hintegralprop}), we finally find the asymptotics for the $\mathcal{H}$-integrals in the $n \to \infty$ regime to be
\begin{align}
\mathcal{H}^{(a,b,c,r,s)}_{n,j} \widesim[3]{n\to \infty} \frac{n^{j+s+b+\frac{D-4}{2}} \left(j+r+\frac{D-4}{2}\right)!}{(\log n)^{\frac{D-2}{2}+r-c}}~.
\end{align}
On plugging the above result into (\ref{term1}, \ref{term2}) and keeping only those terms that have the fastest $n$-growth (i.e., those terms with the largest power of $n$), we have for the actual residue (including the factors contributing from the coefficient $c^D_{n,j}$):

\begin{align}
B^D_{n,j}(q) \widesim[3]{n\to \infty} &\frac{2^{D-3}}{\sqrt{\pi}} \left(j+\frac{D-3}{2}\right) \Gamma\left(\frac{D-3}{2}\right) \frac{1}{n^{D/2} (\log(n))^{\frac{D-2}{2}}} \nonumber \\
&~~~~\times \bigg[(1-(-1)^{n+j}) + (1-q) (-1)^{n+j} \frac{n^2 \log(n)}{2} + \mathcal{O}\big[(1-q)^2;n^2\big]\bigg]~.
\label{asymptoticsfixedj}
\end{align}
We make a couple of interesting observations for the fixed $j$ asymptotics described by the above relation:
\begin{itemize}
    \item We note that (\ref{asymptoticsfixedj}) is consistent with the formula (4.17) derived in \cite{Arkani-Hamed:2022gsa}. For $n+j \in 2\mathbb{Z}$ , the leading order contribution vanishes as expected and we have the first-order contribution to the asymptotics whose values are \textit{positive}:
\begin{align}
B^D_{n,j}(q) \widesim[3]{n\to \infty} (1-q) \frac{2^{D-4}}{\sqrt{\pi}}& \left(j+\frac{D-3}{2}\right) \Gamma\left(\frac{D-3}{2}\right) \frac{1}{(n\log(n))^{\frac{D-4}{2}}} >0~.
\end{align}
For $n+j\in2\mathbb{Z}+1$ , both terms in (\ref{asymptoticsfixedj}) contribute with the first-order term giving a negative contribution. 

    \item We note from (\ref{asymptoticsfixedj}) that the first-order contribution grows \textit{faster} in comparison to the leading order term at a rate $\sim n^2 \log(n)$.
\end{itemize}

\subsubsection{Regge asymptotics}\label{sec:appliregge}
For studying the second type of asymptotics, namely the Regge behavior, we keep $\Delta=n-j$ fixed and take $n\to \infty$. We begin by first writing our basis integrals (\ref{double-contourmasterintegral2}) in a form similar to the one in \cite{Arkani-Hamed:2022gsa}
\begin{equation}
    \oint_{u=0}\frac{du}{2\pi i}\oint_{v=u}\frac{dv}{2\pi i} \frac{1}{(uv)^{\frac{D-2}{2}}u^rv^s}\left(\frac{u^{-1}-v^{-1}}{e^v-e^u}\right)^n\left(u^{-1}-v^{-1}\right)^{-\Delta}(v-u)^c\frac{e^{-bu}e^{(a+b)v}}{(e^v-e^u)^a}~.
\end{equation}
As already mentioned in \cite{Arkani-Hamed:2022gsa}, in the large $n$ limit it is enough to truncate the Taylor expansion of the second term to third order in $(v-u)$ in order to achieve the largest powers of $n$. Hence one has
\begin{equation}
    \frac{u^{-1}-v^{-1}}{e^v-e^u} = \frac{e^{-u}}{u^2} \left(1-\frac{(u+2)(v-u)}{2u}+\frac{(u^2+6u+12)(v-u)^2}{12u^2}\right)+\mathcal{O}(v-u)^3~.
\end{equation}
Then circling around the saddle point at $u=-2$, we can set $u=-2$ wherever doing so gives a non-zero contribution and $u=v$ wherever the poles are absent\footnote{For more details look at section 4.3.2 in \cite{Arkani-Hamed:2022gsa}.}. We get\footnote{The constant of proportionality comes from the gamma functions and the $(-1)^{r+s+1}$ factor in (\ref{integrandpart2}).}
\begin{align}
    &\mathcal{I}^{(a,b,c,r,s)}_{n,n-\Delta} \propto\nonumber\\
    &\oint_{u=0}\frac{du}{2\pi i}\oint_{v=u}\frac{dv}{2\pi i} \frac{1}{u^{D-2+r+s-2\Delta}}\left(1+\frac{(u+2)(u-v)}{4}+\frac{(v-u)^2}{12}\right)^n\frac{1}{(v-u)^{\Delta-c}}\frac{1}{(1-e^{u-v})^a},
\end{align}
to perform the above integral in the $n\to \infty$ we follow similar procedure as in \cite{Arkani-Hamed:2022gsa}. First consider substituting $t=(v-u)\sqrt{n}$ and use the fact that $(1+\frac{a}{n})^{n}\sim e^{a}$ when $n\to\infty$. Now, since $a \in \mathbb{Z}_{\leq 0}$ and the fact that the integral is non-zero if and only if $\Delta>c$, we can simply interchange the integral and take residue w.r.t $u$ to get:
\begin{align*}
    &\mathcal{I}^{(a,b,c,r,s)}_{n,n-\Delta} \propto\frac{n^{-\frac{3\Delta-c-5}{2}+2n+D+r+s}(-1)^{r+s+1}}{(2n-2\Delta+D+r+s-3)!}\oint_{t=0}\frac{dt}{2\pi i} \frac{1}{t^{\Delta-c}}\left(1-\frac{t}{4\sqrt{n}}\right)^{2n}\exp(\frac{1}{2}\sqrt{n}t+\frac{t^2}{12})~.
\end{align*}
Using the fact that in the limit $n\to \infty$
\begin{equation}
    \left(1-\frac{t}{4\sqrt{n}}\right)^{2n}\exp(\frac{1}{2}\sqrt{n}t+\frac{t^2}{12}) = \exp(\frac{t^2}{48})+\mathcal{O}(n^{-\frac{1}{2}})~,
\end{equation}
the remaining $t$-integral is then evaluated as
\begin{equation}
    \oint_{t=0}\frac{1}{2 \pi i}\frac{1}{t^{\Delta-c}}\exp(\frac{t^2}{48}) = \frac{1}{\left(\frac{\Delta-c-1}{2}\right)!}\left(\frac{1}{48}\right)^{\frac{\Delta-c-1}{2}}~,
\end{equation}
which holds for any $\Delta>c$. Then finally using Stirling's approximation as before, we get that the basis integrals in the $n\to \infty$ limit is
\begin{equation}
    \mathcal{I}^{(a,b,c,r,s)}_{n,n-\Delta}\propto \frac{(-1)^{r+s+1}}{\sqrt{\pi n}}\frac{2^{2c+4-D-r-s-2n}}{\left(\frac{\Delta-c-1}{2}\right)!}\left(\frac{n}{3}\right)^{\frac{\Delta-c-1}{2}}e^{2n}~.
\end{equation}
Putting in the constants of proportionality, we finally arrive at 
\begin{equation}
    \mathcal{I}^{(a,b,c,r,s)}_{n,n-\Delta} \sim \frac{2^{2c+5-D-r-s-2n}\sqrt{\pi}}{3^{\frac{\Delta-c-1}{2}}\left(\frac{\Delta-c-1}{2}\right)!}n^{2n+r+s+D-4-\frac{3\Delta+c}{2}}~.
\end{equation}
Plugging this into (\ref{BintermsofI}) and again retaining only those terms that have the fastest $n$-growth, we find for $n+j \in 2\mathbb{Z}+1$
\begin{equation}
    B^{D}_{n,n-\Delta} \widesim[3]{n\to \infty} \frac{2^{\frac{3}{2}-2n}}{3^{\frac{\Delta-1}{2}}}\frac{e^n}{\sqrt{\pi}}\frac{\Gamma\left(\frac{D-3}{2}\right)}{\Gamma\left(\frac{\Delta+1}{2}\right)}n^{\frac{\Delta-D+1}{2}}\left(1+(1-q)\frac{n^3}{24(\Delta+1)}\right)>0~, \hspace{5mm} \forall \Delta \in \{1,\dots,n\}~,
    \label{regge1storder1}
\end{equation}
where as for $n+j \in 2 \mathbb{Z}$, we get
\begin{equation}
    B^{D}_{n,n-\Delta} \widesim[3]{n\to \infty} (1-q)\frac{2^{-\frac{3}{2}-2n}}{3^{\frac{\Delta+1}{2}}}\frac{e^n}{\sqrt{\pi}}\frac{\Gamma\left(\frac{D-3}{2}\right)}{\Gamma\left(\frac{\Delta+1}{2}\right)}\frac{n^{\frac{\Delta-D+7}{2}}}{\Delta+1}>0~, \hspace{5mm} \forall \Delta \in \{1,\dots,n\}~.
    \label{regge1storder2}
\end{equation}
We now make a couple of crucial observations for the Regge asymptotics described by the above relations (\ref{regge1storder1}, \ref{regge1storder2}):
\begin{itemize}
    \item Note the manifest \textit{positivity} at first order for both $n+j \in 2\mathbb{Z}$ and $n+j \in 2\mathbb{Z}+1$, unlike the fixed spin case where we saw this only for $n+j \in 2\mathbb{Z}$. This combined with our numerical analysis away from $q=1$ in section \ref{sec:lvsssr} allows us to conclude positivity for all Regge trajectories for any $q\in[0,1]$ and any spacetime dimension $D\geq4$ in the large $n$ limit. This is consistent with the main result in \cite{Chakravarty:2022vrp} for $D=4$ and $m^2=0$.
    \item Once again, we see that the first-order corrections grow \textit{faster} than the leading Veneziano term at a rate $\sim n^3$.
\end{itemize}

\section{Outlook and discussions}
\label{sec:outlook}
Here we reflect on some of the key results of our paper and discuss the potential scope for future work and some open questions. The main results are summarized as under:
\begin{itemize}
\item Our contour integral representation of the Coon amplitude partial wave coefficients $B^{D}_{n,j}(q)$ given by (\ref{Bqequation}) establishes manifest \textit{positivity} for the leading and sub-leading Regge sectors in arbitrary spacetime dimensions $D$. 
\item Our closed-form, analytic expression of the sub-sub-leading Regge trajectory (\ref{ssrB}) reveals the first signatures of breakdown of positivity for certain critical values of $q, D$ as seen from figures \ref{B3,0plot3D} and \ref{regplot}.
\item We consider a perturbative expansion near the Veneziano amplitude ($q=1$) to get first glimpses of the mathematical structures that one ought to encounter in a double-contour representation for the Coon amplitude coefficients $B^{D}_{n,j}(q)$, in a spirit similar to the Veneziano case \cite{Arkani-Hamed:2022gsa}. In particular, we arrive at a generalized basis of integrals that we expect to show up at all orders in the $(1-q)$ expansion. Studying the properties of these basis integrals, we derive a double-contour representation for the first $\mathcal{O}(1-q)$ contribution to the coefficients $B^{D}_{n,j}(q)$. This expression finds applications in studying the $n \to \infty$ asymptotic limit of Regge trajectories.
\item For $q$ not very close to 1, our numerical analysis from section \ref{sec:lvsssr} suggests the positivity of \textit{all} Regge trajectories based on the manifest positivity of the leading Regge coefficient. For $q \sim 1$ the numerical analysis is unreliable; however, our analytic expressions (\ref{regge1storder1}, \ref{regge1storder2}) assert the positivity of such Regge trajectories in the asymptotic regime for arbitrary $D$. A similar discussion also holds that argues for the positivity of fixed $j$ asymptotics based on the numerical analysis presented in appendix \ref{app: numfixjasymp} and the analytic expression (\ref{asymptoticsfixedj}). \end{itemize}

\begin{figure}[ht]
    \centering
     \subfloat[\centering ]{{\includegraphics[scale=1.60]{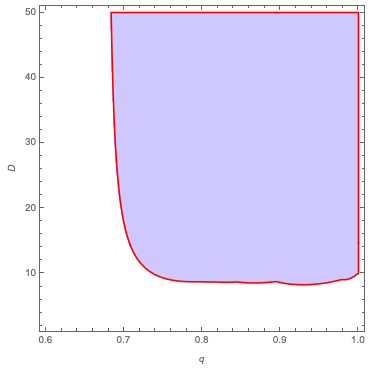}}} %
     \subfloat[\centering ]{{\includegraphics[scale=1.60]{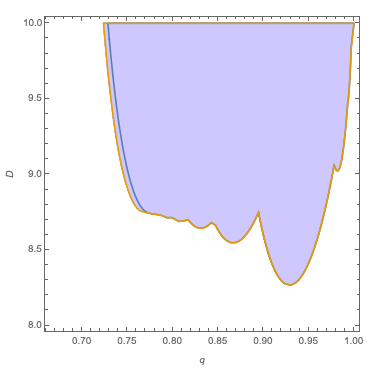}}} %
    \caption{Plots indicating the unitarity envelope for the Coon amplitude, computed analytically from $B_{n,0}^{D}(q)$ with a cutoff $n_{\rm max}$. The Coon amplitude violates unitarity in the shaded region of $(q,D)$ parameter space. Plot (a), with $n_{\rm max} = 8$, already agrees well with the numerically computed unitarity envelope depicted by the red-dotted curve ($m^2=0$ case) in Fig.~3 of \cite{Figueroa:2022onw}. Plot (b) is a closeup of the region $q \in [2/3,1]$, $D \in [8,10]$, showing the small change as we increase the cutoff from $n_{\rm max}=10$ (dark blue) to $n_{\rm max}=13$ (gold).  We speculate that the $q \gtrsim 0.77$ part of the boundary curve is already stable (which in particular would mean we know its shape analytically), while the excluded region grows slowly to the left as the cutoff is increased.} \label{unitaritycurve}
\end{figure}

We know from the work of~\cite{Figueroa:2022onw} that the Coon amplitude is unitary in all dimensions for $q \le \frac{2}{3}$ and that there exists a unitary envelope for $q \in (\frac{2}{3},1)$ (which our analysis confirms). On the other hand, figure \ref{regplot}(b) establishes a minimal region in $(q,D)$ parameter space in which the Coon amplitude is definitely not unitary. It remains an open problem to analytically prove or disprove unitarity elsewhere within $q \in [0,1)$, $D \in (4,10]$.  Our results (based on numerical experimentation for $q$ not close to 1, and on analytic analysis for $q$ near 1) suggest that the region in $(q,D)$ space is excluded by demanding that $B_{n+m,m}^{(D)}(q) \ge 0$ always sits \emph{inside} the region excluded by demanding that $B_{n,0}^{(D)}(q) \ge 0$. If true, then the boundary of the excluded region in $(q,D)$ space can be determined entirely from the latter. We plot the regions excluded by demanding that $B_{n,0}^{(D)}(q) \ge 0$ for several values of $n$ in Fig.~\ref{unitaritycurve}.  Tantalizingly, these plots suggest (but again, do not prove) that the regions excluded by considering larger values of $n$ lie at smaller values of $q$ than those at smaller values of $n$. If this is true, it means that the analytic shape of the boundary curve for any finite value of $q$ is, in principle, analytically determinable from explicit knowledge of $B_{n,0}^{(D)}(q)$.

A general proof of unitarity in certain regions may be attainable either from an amplitudes perspective or a $q$-string analog of the \textit{no-ghost} theorem. The former approach has been recently pioneered in \cite{Arkani-Hamed:2022gsa} for the Veneziano amplitude, which served as our main source of inspiration to address this question for Coon. It would be very interesting to extend our analysis to the new classes of four-point amplitudes recently constructed in \cite{Cheung:2022mkw,Geiser:2022exp}. Of course, it would also be very interesting to explore higher-point amplitudes, which we expect will lead to much more stringent constraints on unitarity than four-point amplitudes do.

\acknowledgments
We are grateful to Adam Ball, Luke Lippstreu, Andrzej Pokraka, Lecheng Ren and Akshay Yelleshpur Srikant for useful comments and discussion.
This work was supported in part by the US Department of Energy under contract {DE}-{SC}0010010 Task F and by Simons Investigator Award \#376208.

\appendix

\section{A primer on \texorpdfstring{$q$}{q}-analysis}
\label{sec:q-calculusprimer}
In this appendix, we present some important $q$-theory machinery that is central to our manipulations in passing from Veneziano to the Coon amplitude. We refer the interested reader to standard texts on the subject \cite{kac2002quantum, ernst2012comprehensive} for extensive details. We first define the $q$-integers $[n]_q$ (also known as the $q$-bracket of ordinary non-negative integers $n$) by\footnote{It is useful to note that this particular choice of defining the $q$-integers (among many other possible options) provides a natural starting point to define $q$-analogs of other associated quantities like the $q$-factorial, $q$-Pochhammer and so on.}:
\begin{equation}
    [n]_q = \frac{1-q^n}{1-q}~,
\label{q-integers}
\end{equation}
where $q$ is a fixed real number $0<q<1$. As with all $q$-deformed objects, one recovers the original identity or expression in the limit $q \to 1$. For the $q$-integers defined above, one can easily check that $\lim_{q \to 1} [n]_q = n$. With that, the definition of the $q$-factorial arises naturally:
\begin{align}
[n]_q! = [1]_q \cdot [2]_q \cdots [n]_q = \frac{(q;q)_n}{(1-q)^n}~,
\label{q-factorial}
\end{align}
where in the final equality we have introduced the $q$-Pochhammer symbol defined by 
\begin{align}
(a;q)_n = \prod_{j=0}^{n-1} (1-aq^j)~.  
\label{q-pochhammer}
\end{align}
This is a $q$-analog of the ordinary Pochhammer symbol $(x)_n$ in the sense that $\lim_{q\to 1} \frac{(q^x;q)_n}{(1-q)^n} = (x)_n$ $\forall x \in \mathbb{R}$. We note that the $q$-Pochhammer symbol has an infinite product representation (unlike its ordinary counterpart):
\begin{align}
(a,q)_{\infty} = \prod_{j=0}^{\infty} (1-aq^j)~.
\end{align}
However, for most of the analysis in our paper, we shall use the $q$-Pochhammer notation established in \cite{kac}:
\begin{align}
    (a+b)^n_q &= \prod_{j=0}^{n-1}(a+q^jb)~, \label{infprodqpoch1} \\
    (1+a)^{\infty}_q &= \prod_{j=0}^{\infty}(1+q^ja)~, \label{infprodqpoch2}\\
    (1+a)^{t}_q &= \frac{(1+a)^{\infty}_q}{(1+q^ta)^{\infty}_q}~, \hspace{5mm} \forall t\in \mathbb{C}~. \label{infprodqpoch3}
\end{align}
We note that the infinite product (\ref{infprodqpoch1}) is convergent since $0<q<1$.   

\subsection{Elements of \texorpdfstring{$q$}{q}-calculus}
We introduce some basic notions of $q$-calculus that prove handy in computing derivatives and integrals of $q$-deformed quantities. We first introduce the $q$-derivative given by
\begin{equation}
    D_{q}f(x) = \frac{f(qx)-f(x)}{(q-1)x}~.
    \label{q-derivative}
\end{equation}
Consequently, we define the notion of the $q$-antiderivative given by
\begin{equation}
    \int f(x) d_qx = (1-q) x \sum_{j=0}^{\infty} q^{j} f(q^jx)~,
    \label{jacksonint}
\end{equation}
with the series being called the \textit{Jackson integral} of $f(x)$. The two definitions (\ref{q-derivative}) and (\ref{jacksonint}) combine to give the fundamental theorem of $q$-calculus.  
\paragraph{Theorem A.1.1} (\textit{Fundamental theorem of $q$-calculus}) If $F(x)$ is an antiderivative of $f(x)$ and $F(x)$ is continuous at $x=0$, we have
\begin{equation}
    \int_a^b f(x) d_qx = F(b)-F(a)~,
    \label{fundtheoremq-calc}
\end{equation}
where $0 \leq a < b \leq \infty$. 
\paragraph{Corollary A.1.1} If $f^{\prime}(x)$ exists in a neighbourhood of $x=0$ and is continuous at $x=0$, where $f^{\prime}(x)$ denotes the ordinary derivative of $f(x)$, we have
\begin{equation}
\int_{a}^{b} D_qf(x) d_qx = f(b)-f(a)~.
\label{corollary}
\end{equation}
We list some of the important properties of these $q$-derivatives and Jackson integrals as follows:
\begin{align}
    \int_{a}^{b}f(x)d_qx := \int_{0}^{b}f(x)d_qx-\int_{0}^{a}f(x)d_qx~.
\end{align}
We also have a natural formula of $q$-integration by parts: 
\begin{equation}
    \int_{0}^{a}g(x)D_qf(x)d_qx = f(x)g(x)\Big|_0^a-\int_{0}^{a}f(qx)D_qg(x)d_qx~,
\end{equation}
which follows from the $q$-Leibniz rule\footnote{We note that by symmetry, we can interchange $f$ and $g$, and obtain another equivalent version of the $q$-Leibniz rule $D_q(f(x)g(x)) = f(x)D_qg(x)+g(qx)D_qf(x)$.}
\begin{equation}
    D_q(f(x)g(x)) = f(qx)D_qg(x) + g(x)D_qf(x)~.
    \label{leibnizrule}
\end{equation}
However, one should note that there does not exist a general chain rule for $q$-derivatives. An exception is the derivative of a function of the form $f(u(x))$, where $u(x) = \alpha x^{\beta}$ with $\alpha, \beta$ being constants, for which we have
\begin{align}
    D_qf(u(x)) &= (D_{q^{\beta}}f)(u(x)) \cdot D_{q} u(x) \\
    \int_{u(a)}^{u(b)} f(u)d_qu &= \int_{a}^{b} f(u(x)) D_{q^{1/\beta}}u(x) d_{q^{1/\beta}}x.
\end{align}
We list below a few known examples of basic $q$-derivatives that follow immediately from the $q$-Leibniz rule and the definitions (\ref{infprodqpoch1})-(\ref{infprodqpoch3}):
\begin{align}
    D_qx^{t} &= [t]_qx^{t-1} \label{derivativeex1}~, \\
    D_{q}(Ax+b)^{n}_q &= [n]_qA(Ax+b)^{n-1}_q \label{derivativeex2}~, \\
    D_{q}(a+Bx)^{n}_q &= [n]_qB(a+Bqx)^{n-1}_q~.
\end{align}
For an extensive list of results, we refer the reader to \cite{kac}.

\subsection{Lambert series}
We briefly review the Lambert series which comes in handy when dealing with the Taylor expansion of the contour integral (\ref{Bqequation}) around $q=1$, as shown later in Appendix \ref{sec:taylorexpappen}. For any $s\in \mathbb{C}$ and $z>0$, the Lambert series is defined as the following convergent sum \cite{banerjee2016lambert, banerjeeandblake}:
\begin{equation}
    \mathcal{L}_q(s,z) = \sum_{k=1}^{\infty}\frac{k^sq^{kz}}{1-q^k}~,
    \label{lambertdefn}
\end{equation}
where $q\in[0,1)$. The Lambert series is closely related to the $q$-Pochhammer symbol via the identity:
\begin{equation}
    \log{(1-q^z)_q^{\infty}} = -\mathcal{L}_q(-1,z)\label{lamtopoch}~.
\end{equation}
To derive this rather important identity, we first consider
\begin{align}
    \log(1-q^z)_q^{\infty} &= \log \prod_{n=0}^{\infty}(1-q^{n+z})= \sum_{n=0}^{\infty}\log(1-q^{n+z}) \nonumber \\
    &= -\sum_{n=0}^{\infty}\sum_{k=1}^{\infty}\frac{q^{k(n+z)}}{k}~,
\end{align}
where the second equality follows from Taylor expanding the logarithm around $z=0$. Now since we assume that $z>0$ and $q\in [0,1)$, the two sums converge and we can therefore interchange them to sum over the $n$ series first
\begin{align}
    \log(1-q^z)_q^{\infty} &= -\sum_{k=1}^{\infty}\frac{q^{kz}}{k}\sum_{n=0}^{\infty}q^{kn}\\
    &= -\sum_{k=1}^{\infty}\frac{1}{k}\frac{q^{kz}}{1-q^k}=-\mathcal{L}_q(-1,z)
\end{align}
to get the desired result. In going from the first to the second line, we have summed over the geometric series for $q^{k}\in [0,1)$, and in the final expression we employed the definition of the Lambert series (\ref{lambertdefn}). 

\subsection{\texorpdfstring{$q$}{q}-gamma and \texorpdfstring{$q$}{q}-beta functions}
\label{sec:q-gammabeta}
We introduce the $q$-analogs of the standard gamma and Euler-beta functions whose various representations are central to our analysis. We start with reviewing the $q$-deformation of the simplest transcendental function in analysis, the exponential, for which there exists two $q$-analogs, namely: 
\begin{align}
    E^{x}_q &= \sum_{n=0}^{\infty}q^{n(n-1)/2}\frac{x^n}{[n]_q!} = (1+(1-q)x)^{\infty}_q~,\\
    e^{x}_q &= \sum_{n=0}^{\infty}\frac{x^n}{[n]_q!} = \frac{1}{(1-(1-q)x)^{\infty}_q}~.
\end{align}
The two definitions are unfortunately not equivalent but instead inverses of each other: 
\begin{equation}
    e^{x}_qE^{-x}_q = E^{x}_qe^{-x}_q = 1~,
\end{equation}
with their $q$-derivatives being
\begin{equation}
    D_qe^{x}_q = e^{x}_q~, \qquad D_qE^{x}_q = E^{qx}_q~.
\end{equation}
Similarly, there are two equivalent definitions of the trigonometric functions which we will not review here (one can refer to \cite{kac, kac2002quantum} for more details). 

We now define the $q$-gamma function which admits the following infinite product representation: 
\begin{equation}
    \Gamma_{q}(t) := \frac{(1-q)^{t-1}_q}{(1-q)^{t-1}}= \frac{\prod_{n=0}^{\infty}1-q^{t-1+n}}{(1-q)^{t-1}}~, \hspace{5mm} t>0~.
\end{equation}
It also has several integral representations one of which is 
\begin{equation}
    \Gamma_{q}(t) = \int_{0}^{1/(1-q)} x^{t-1}E^{-qx}_qd_qx~.
\end{equation}
As in the standard case, one can also define the $q$-beta function in terms of these $q$-gamma functions as follows 
\begin{equation}
    B_q(s,t) := \frac{\Gamma_q(t)\Gamma_q(s)}{\Gamma_q(t+s)}~,
\end{equation}
which also has a useful integral representation of the form
\begin{equation}
    B_{q}(s,t) = \int_{0}^{1}x^{t-1}(1-qx)^{s-1}_qd_qx~.
    \label{integralq-beta}
~\end{equation}
\section{Details of the derivation for the \texorpdfstring{$\mathcal{O}(1-q)$}{O(1-q)} double-contour representation} \label{sec:taylorexpappen}
\subsection{Relevant Taylor expansions}
\label{sec:relevanttaylorexp}
In this appendix we fill in the details regarding crucial Taylor expansions in section \ref{sec: smallqdeform}. To consider small deformations around the Veneziano amplitude, we Taylor expand our integral result (\ref{Bqequation}) around $q=1$ up to first order. To that end, we manipulate the $q$-deformed quantities one by one. At first, we deal with the coefficients $c_{n,j}^D(q)$ for which we get
\begin{equation}
    c_{n,j}^D(q) = c^{D}_{n,j}+c^{D}_{n,j}\left(\frac{D+j-2}{2}(n-1)-n\right)(1-q)+\mathcal{O}(1-q)^2~.
    \label{cexpansion}
\end{equation}
Moreover, we note that $(1-(1-q)t)^{n-1} = 1-(n-1)(1-q)t+\mathcal{O}(1-q)^2$. Now, the slightly trickier part is Taylor expanding the integration range, the measure and the integrand $(t+[n]_q)^{\frac{D-4}{2}+j}$, we get:
\begin{align}
    &\int_{-[n]_q}^{0} dt (-t(t+[n]_q))^{\frac{D-4}{2}+j}(1-z)^{-t-1} \nonumber \\ 
    &= \int_{-n\left(1-\frac{n-1}{2}(1-q)\right)}^{0} dt \left(-t(t+n)\right)^{\frac{D-4}{2}+j}\left(1-\left(\frac{D-4}{2}+j\right)\frac{n(n-1)}{2(t+n)}(1-q)\right)(1-z)^{-t-1}~\nonumber\\
    &~~~~~~~~~~~~~~~~~~~+ \mathcal{O}(1-q)^2~. \label{someequation}
\end{align}
Considering the substitution $t \to t(1-\frac{n-1}{2}(1-q))$ we finally get 
\begin{align}
    \int_{-[n]_q}^{0} dt &(-t(t+[n]_q))^{\frac{D-4}{2}+j}(1-z)^{-1-t}(-\log(1-z))^{j}\nonumber\\
    &= \int_{-n}^{0}dt\left(-t(t+n)\right)^{\frac{D-4}{2}+j} (1-z)^{-1-t}(-\log(1-z))^{j}\nonumber\\
    &~~~~-(1-q)\bigg[\frac{n-1}{2}\left(\frac{D-2}{2}+j\right)\int_{-n}^{0}dt\left(-t(t+n)\right)^{\frac{D-4}{2}+j} (1-z)^{-1-t}\nonumber\\
    &~~~~-\frac{n-1}{2}\log(1-z)\int_{-n}^{0}dt(-t)^{\frac{D-2}{2}+j}(t+n)^{\frac{D-4}{2}+j}(1-z)^{-1-t}\nonumber\\
    &~~~~+\left(\frac{D-4}{2}+j\right)\frac{n(n-1)}{2}\int_{-n}^{0}dt (-t)^{\frac{D-4}{2}+j}(t+n)^{\frac{D-6}{2}+j}(1-z)^{-1-t}\nonumber\\
    &~~~~+\frac{n-1}{2}\left(\frac{D-4}{2}+j\right)\int_{-n}^{0}dt (-t)^{\frac{D-2}{2}+j}(t+n)^{\frac{D-6}{2}+j}(1-z)^{-1-t}\bigg](-\log(1-z))^{j}~. 
    \label{integrandpart1}
\end{align}
Finally we Taylor expand the $(1-qz)_q^{-\alpha_q(t)-1}$ term inside the derivative in equation (\ref{Bqequation}), but first note 
\begin{align}
    (1-qz)^{x}_q &\equiv \frac{(1-qz)^{\infty}_q }{(1-q^{1+x}z)_q^{\infty}} = \exp\left\{\mathcal{L}_q\left(-1,\frac{\log{zq^{1+x}}}{\log{q}}\right)-\mathcal{L}_q\left(-1,\frac{\log{zq}}{\log{q}}\right)\right\}~,
\end{align}
where we made use of the crucial identity derived in the previous section given by equation (\ref{lamtopoch}). For $x= -\alpha_q(t)-1$, we have:
\begin{equation*}
    (1-qz)^{-\alpha_q(t)-1}_q = \exp\left\{\sum_{k=1}^{\infty}\frac{z^k(1-(1-q)t)^{-k}}{k(1-q^k)}-\sum_{k=1}^{\infty}\frac{z^kq^{k}}{k(1-q^k)}\right\}~.
\end{equation*}
Taylor expanding to first order around $q=1$ we get
\begin{align}
   (1-qz)^{-\alpha_q(t)-1}_q &= \exp\left\{(1+t)\sum_{k=1}^{\infty}\frac{z^k}{k}+\frac{t^2}{2}(1-q)\sum_{k=1}^{\infty}\frac{k+1}{k}z^k\right\} \nonumber \\
   &= (1-z)^{-1-t}\left[1+\left(\frac{1-q}{2}\right)\left(\frac{z}{1-z}-\log(1-z)\right)t^2\right]+\mathcal{O}(1-q)^2\label{eq:taylorzq}.
\end{align}
Putting this back into the integral we get 
\begin{align}
    \int_{-n}^{0} dt (-t(t+n))^{\frac{D-4}{2}+j}\partial_{t}^{j}\left\{(1-z)^{-1-t}\left[1-(n-1)(1-q)t+(1-q)\left(\frac{z}{1-z}-\log(1-z)\right)\frac{t^2}{2}\right]\right\}~,
\end{align}
and, differentiating w.r.t $t$, we get
\begin{align}
     &    \int_{-n}^{0} dt~(-t(t+n))^{\frac{D-4}{2}+j}\partial_{t}^{j}\left\{(1-z)^{-1-t}\left[1-(n-1)(1-q)t+(1-q)\left(\frac{z}{1-z}-\log(1-z)\right)\frac{t^2}{2}\right]\right\} \nonumber \\ 
     &= (-\log(1-z))^j\int_{-n}^{0}dt~(-t(t+n))^{\frac{D-4}{2}+j}(1-z)^{-1-t} \nonumber  \\
     &-(1-q)\Bigg\{(n-1)\bigg[j(-\log(1-z))^{j-1}\int_{-n}^{0}dt~(-t(t+n))^{\frac{D-4}{2}+j}(1-z)^{-1-t} \nonumber \\
     &\qquad \qquad \qquad \qquad \qquad \qquad -(-\log(1-z))^{j}\int_{-n}^{0}dt~ (-t)^{\frac{D-2}{2}+j}(t+n)^{\frac{D-4}{2}+j}(1-z)^{-1-t}\bigg] \nonumber  \\
     &~~~~~~~~~~~-\frac{1}{2}\bigg(\frac{z}{1-z}-\log(1-z)\bigg)\bigg[j(j-1)(-\log(1-z))^{j-2}\int_{-n}^{0} dt~(-t(t+n))^{\frac{D-4}{2}+j} (1-z)^{-1-t} \nonumber \\
     &~~~~~~~~~\qquad \qquad \qquad \qquad \qquad \qquad  -2j(-\log(1-z))^{j-1}\int_{-n}^{0}dt~(-t)^{\frac{D-2}{2}+j}(t+n)^{\frac{D-4}{2}+j}(1-z)^{-1-t} \nonumber  \\
     &\qquad \qquad \qquad  ~~~~~~~~~ +\left(-\log(1-z)\right)^{j}\int_{-n}^{0} dt~(-t)^{\frac{D}{2}+j}(t+n)^{\frac{D-4}{2}+j}(1-z)^{-1-t}\bigg]\Bigg\}~. \label{integrandpart2}
\end{align}
Combining (\ref{cexpansion}), (\ref{integrandpart1}) and (\ref{integrandpart2}), we have the $\mathcal{O}(1-q)$ Taylor expansion of the $H_q(z)$ integral.

\subsection{Deriving double contour integral representation of \texorpdfstring{$\mathcal{H}$}{H}-integrals}

We now derive the double contour integral representation for the $\mathcal{H}$-basis integrals we introduced in section \ref{sec: smallqdeform}. The derivation is more or less a generalization of the one presented in \cite{Arkani-Hamed:2022gsa}. First consider the $H$-integral already defined in section \ref{sec: smallqdeform}
\begin{align}
H_{n,j}^{(r,s)}(t^{\prime}) = \int^{t^\prime} dt \frac{(-t)^{\frac{D-4}{2}+j+r} (t+n)^{\frac{D-4}{2}+j+s}}{(1-z)^{t+1}}~,    
\label{h-integral}
\end{align}
where $r,s \in \mathbb{Z}$. Substituting $u = \log(1-z)$, we have
\begin{align}
H_{n,j}^{(r,s)}(t^{\prime})
&=e^{nu} \int^{t^\prime} du\, e^{-(t+1+n)u} (-t)^{\frac{D-4}{2}+j+r} (t+n)^{\frac{D-4}{2}+j+s} \nonumber \\
&= (-1)^{s+r} e^{(n-1)u} \partial_u^{\frac{D-4}{2}+j+s} \int^{t^\prime} du \, e^{-(t+n)u} t^{\frac{D-4}{2}+j+r}~, \label{h-integral2}
\end{align}
we try to realize the integrand as a total derivative by observing the following fact:
\begin{align}
\partial_t\left(\frac{e^{-(t+n)u}}{u^{j+r+\frac{D-2}{2}}} \sum_{k=0}^{j+r+\frac{D-4}{2}} \frac{(tu)^k}{k!}\right) &= \frac{e^{-(t+n)u}}{u^{j+r+\frac{D-4}{2}}} \left(-\sum_{k=0}^{j+r+\frac{D-4}{2}} \frac{(tu)^k}{k!} + \sum_{k=0}^{j+r+\frac{D-6}{2}} \frac{(tu)^k}{k!}\right) \nonumber \\
&= - \frac{e^{-(t+n)u} t^{\frac{D-4}{2}+j+r}}{(j+r+\frac{D-4}{2})!}~.
\end{align}
This implies that the integrand can be rewritten as a total derivative
\begin{align}
e^{-(t+n)u} t^{\frac{D-4}{2}+j+r} = \left(j+r+\frac{D-4}{2}\right)! \, \partial_t\left(\frac{e^{-(t+n)u}}{u^{\frac{D-2}{2}+j+r}} \sum_{k=0}^{j+r+\frac{D-4}{2}} \frac{(tu)^k}{k!}\right)~,
\end{align}
which means that the $H$-integral can be written as
\begin{align}
H_{n,j}^{(r,s)}(t^{\prime}) = (-1)^{s+r+1} e^{(n-1)u} \left(j+r+\frac{D-4}{2}\right)! \partial_u^{j+s+\frac{D-4}{2}} \left(\frac{e^{-(t^{\prime}+n)u}}{u^{j+r+\frac{D-2}{2}}} \sum_{k=0}^{j+r+\frac{D-4}{2}} \frac{(t^{\prime}u)^k}{k!}\right)~.
\label{h-final}
\end{align}
We now define the generalised basis for our double contour integral representation already introduced section \ref{sec: smallqdeform}:
\begin{equation}
    \mathcal{H}^{(a,b,c,r,s)}_{n,j}(t') := \oint \frac{dz}{2\pi i}\frac{1}{z^{n+a}}\frac{1}{(1-z)^b}(-\log(1-z))^{j+c}H^{(r,s)}_{n,j}(t') \label{basis}~.
\end{equation}
We note that from the Taylor expansions (\ref{someequation}), (\ref{eq:taylorzq}) and (\ref{integrandpart2}), it is clear that these are the only integrals that show up at any order in the perturbation series and thus serve as a complete basis for a Taylor expansion around $q=1$ of the partial wave coefficients. 

Moving on, let us first demonstrate how one can relate $\mathcal{H}^{(a,b,c,r,s)}_{n,j}(-n)$ with $\mathcal{H}^{(a,b,c,r,s)}_{n,j}(0)$ of which we find a double contour integral representation later on. Consider the substitution $z \to z/(z-1)$ and $t \to -t-n$ then (\ref{h-integral}) at $t' = -n$ becomes
\begin{align}
    H^{(r,s)}_{n,j}(-n) &= \int^{-n}\frac{dt}{(1-z)^{1+t}}(-t)^{\frac{D-4}{2}+j+r}(t+n)^{\frac{D-4}{2}+j+s}~, \nonumber \\
&= -\int^{0}dt(1-z)^{1-t-n}(-t)^{\frac{D-4}{2}+j+s}(t+n)^{\frac{D-4}{2}+j+r}~, \nonumber \\
    &= -(1-z)^{2-n}H^{(s,r)}_{n,j}(0)~.
\end{align}
Then (\ref{basis}) at $t' = -n$ becomes 
\begin{align}
    \mathcal{H}^{(a,b,c,r,s)}_{n,j}(-n) &= (-1)^{n+j+a+c}\oint \frac{dz}{z^{n+a}}(1-z)^{a+b}(-\log(1-z))^{j+c}H^{(s,r)}_{n,j}(0)\nonumber\\
    &= (-1)^{n+j+a+c} \mathcal{H}^{(a,-a-b,c,s,r)}_{n,j}(0) \label{hbasisid}~.
\end{align}
Now combining the above results with our observation in (\ref{h-final}) and considering the substitution $u = \log(1-z)$ we arrive at 
\begin{align}
    \mathcal{H}^{(a,b,c,r,s)}_{n,j}(0) &= (-1)^{j+c+r+s}\left(j+\frac{D-4}{2}+r\right)! \nonumber \\
    &~~~~~~~~~~~\times\oint \frac{du}{2\pi i}\frac{1}{(1-e^u)^{n+a}}e^{(n-b)u}u^{j+c}\partial_u^{j+\frac{D-4}{2}+s}\left(\frac{e^{-nu}}{u^{\frac{D-2}{2}}+j+r}\right)~.
\end{align}
Performing an integration by parts $j+s+\frac{D-4}{2}$ times in $u$,  the above integral is recast as
\begin{align}
 \mathcal{H}^{(a,b,c,r,s)}_{n,j}(0) = (-1)^{c+r+\frac{D-4}{2}} \left(j+r+\frac{D-2}{2}\right)!\oint_{u=0} \frac{du}{2 \pi i} \frac{e^{-nu}}{u^{\frac{D-2}{2}+j+r}} \partial_u^{\frac{D-2}{2}+j+s}\left(\frac{e^{(n-b)u}u^{j+c}}{(1-e^u)^{n+a}}\right)~.     
\end{align}
This is followed by rewriting the derivative as a residue integral using the identity
\begin{align}
\partial_u^J f(u) = (-1)^J \oint_{v=0} \frac{dv}{2 \pi i} \frac{J!}{v^{J+1}} f(u-v)~,    
\end{align}
with $J=j+s+\frac{D-4}{2}$. This gives us
\begin{align}
 \mathcal{H}^{(a,b,c,r,s)}_{n,j}(0) &= (-1)^{j+c+r+s} \Gamma\left(j+r+\frac{D-2}{2}\right) \Gamma\left(j+s+\frac{D-2}{2}\right) \nonumber \\
 &~~~~~~\times \oint_{u=0} \frac{du}{2 \pi i} \frac{e^{-nu}}{u^{\frac{D-2}{2}+j+r}}\oint_{v=0} \frac{dv}{2 \pi i} \frac{1}{v^{\frac{D-2}{2}+j+s}} \frac{e^{(n-b)(u-v)}(u-v)^{j+c}}{(1-e^{u-v})^{n+a}}~.
\end{align}
Rearranging the terms, we finally obtain a double-contour integral representation for the $\mathcal{H}$-integral as follows
\begin{align}
\mathcal{H}^{(a,b,c,r,s)}_{n,j}(0) &= (-1)^{r+s} \Gamma\left(j+r+\frac{D-2}{2}\right) \Gamma\left(j+s+\frac{D-2}{2}\right) \nonumber \\
&~~~~~~\times \oint_{u=0} \frac{du}{2 \pi i} \oint_{v=0} \frac{dv}{2 \pi i} \frac{(v-u)^{j+c} e^{-bu} e^{(a+b)v}}{(uv)^{\frac{D-2}{2}+j} u^r v^s (e^v-e^u)^{n+a}}~. \label{double-contourmasterintegral}
\end{align}

\section{Numerics for fixed \texorpdfstring{$j$}{j} asymptotics}\label{app: numfixjasymp}
In this appendix we present certain graphical plots of the coefficients $B^D_{n,j}(q)$ for certain \textit{fixed} values of $j$, that are central to the discussion concerning their positivity in the asymptotic regime ($n \to \infty$) in section \ref{sec:outlook}.  

\begin{figure}[ht]
    \centering
    \subfloat[\centering]{{\includegraphics[scale=0.372]{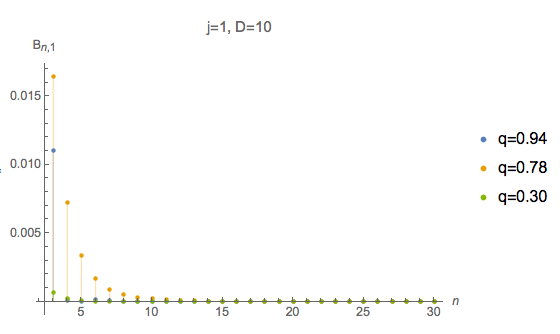}}}%
    \qquad
    \subfloat[\centering]{{\includegraphics[scale=0.372]{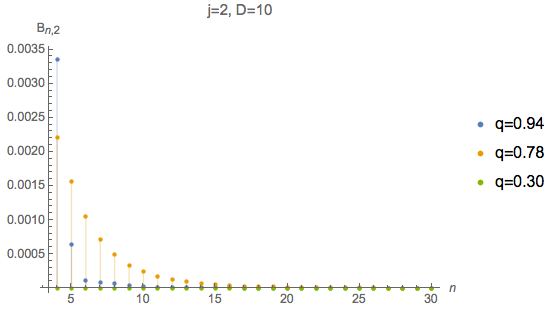}}}%
    \qquad
    \subfloat[\centering]{{\includegraphics[scale=0.372]{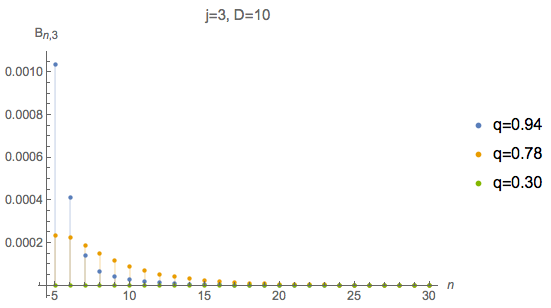}}}%
    \qquad
    \subfloat[\centering]{{\includegraphics[scale=0.372]{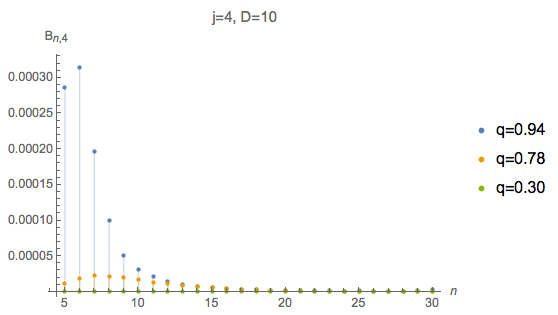}}}%
    \caption{This plots the coefficients $B^{D}_{n,j}(q)$ for fixed values of $j \in \{1,2,3,4\}$ and $D=10$.}
      \label{fig:fixedjasymptoticspositivity}
\end{figure}

We observe that these coefficients die off rapidly for $n \gg 1$, $q\ll 1$. It is also clear that the coefficients follow a monotonically decreasing sequence with respect to spin $j$. In the special case of $D=4$, the noted pattern is consistent with the graphical analysis presented in \cite{Chakravarty:2022vrp}. However, we also see from figure \ref{fig:fixedjasymptoticspositivity} that closer to $q=1$, the coefficients die off less and less rapidly as $j$ increases. Moreover, the pattern of the coefficients around $q=1$ is highly unpredictable as it is non-trivial to analyze their convergent behaviour for larger values of $n$. One can check graphically that this behaviour stabilizes for sufficiently small values of $(1-q)$ and significantly large values of $n$, such that one recovers a decaying pattern similar to the plots in figure \ref{fig:fixedjasymptoticspositivity} which is in qualitative agreement with the expression (\ref{asymptoticsfixedj}).  

\newpage
\bibliography{main}
\bibliographystyle{JHEP}

\end{document}